\documentclass[a4paper,11pt]{article}
\pdfoutput=1
\usepackage{jheppub}
\usepackage{xcolor,bm,comment,braket,mathtools}
\usepackage[T1]{fontenc}
\usepackage{comment}
\usepackage{graphics}

\newcommand{\nn}{\nonumber}
\def\p{\partial}

\def\ul{\underline}

\newcommand {\beq}{\begin{eqnarray}}
\newcommand {\eeqq}{\end{eqnarray}}

\newcommand {\tr}{{\rm tr}\,}

\def\hsymbu#1{\smash{\lower2.5ex\hbox{\huge$#1$}}}

\makeatletter
\newcommand{\doublehat}[1]{%
\begingroup%
  \let\macc@kerna\z@%
  \let\macc@kernb\z@%
  \let\macc@nucleus\@empty%
  \hat{\raisebox{.32ex}{\vphantom{\ensuremath{#1}}}\smash{\hat{#1}}}%
\endgroup%
}
\makeatother

\makeatletter
\@addtoreset{equation}{section}

\renewcommand{\thefootnote}{\fnsymbol{footnote}}
\makeatother

\PassOptionsToPackage{unicode}{hyperref}
\PassOptionsToPackage{naturalnames}{hyperref}

\begin{document}
\thispagestyle{empty}

\title{
\centering{
Ghost-Free Stable Minkowski Vacua \\ in Lovelock Compactifications \\ 
on Irreducible Symmetric Spaces}}

\affiliation[1]{Research and Education Center for Natural Sciences, Keio University, Kanagawa 223-8521, Japan}

\affiliation[2]{Department of Physics, 
Keio University, 4-1-1 Hiyoshi, Kanagawa 223-8521, Japan}

\author[1,2]{Keisuke Ohashi}
\emailAdd{keisuke084@gmail.com}


\abstract{
We study the compactification of higher-dimensional Lovelock gravity on compact
irreducible symmetric spaces, focusing on conditions under which a physically
healthy four-dimensional Minkowski vacuum exists.
We show that when the internal dimension is five or less, or when the theory is
restricted to the Einstein–Gauss–Bonnet sector, the four-dimensional graviton
(tensor sector) is necessarily a ghost.
Inclusion of the cubic Lovelock term removes this ghost instability; however,
the resulting Minkowski vacuum is generically only metastable, being accompanied
by energetically favored Anti–de Sitter vacua.
While such metastability cannot be avoided for spherical internal spaces,
we identify an infinite class of higher-rank symmetric spaces where the true
vacuum can be pushed to infinity in moduli space, thereby realizing 
genuinely stable and ghost-free Minkowski vacua at the level of the four-dimensional effective
theory.
To support these conclusions, we explicitly compute Lovelock terms up to cubic
order on these spaces, confirming a universal log-convexity among the linear,
quadratic, and cubic invariants, which plays a central role in our analysis.
}

\maketitle

\setcounter{page}{1}
\setcounter{footnote}{0}
\renewcommand{\thefootnote}{\arabic{footnote}}

\setcounter{tocdepth}{3}
\section{Introduction}
The problem of spacetime singularities in General Relativity (GR) strongly
suggests that Einstein’s theory should be regarded as an effective field theory
(EFT) of a more fundamental, ultraviolet-complete framework.
From this perspective, it is natural to expect that higher-curvature terms
constructed from the Riemann tensor provide the leading corrections to the
short-distance behavior of gravity.
Among various extensions, Lovelock gravity \cite{Lovelock1971,Lovelock1972}
occupies a distinguished position: it is the most general class of gravitational
theories whose equations of motion remain second order, thereby avoiding the
appearance of Ostrogradsky ghosts that typically afflict higher-derivative
theories.
 Beyond its mathematical elegance, Lovelock gravity is well-motivated by string theory; for instance, the Gauss-Bonnet term emerges in the low-energy effective action of the heterotic string \cite{ZWIEBACH1985315}.

In four spacetime dimensions, the Gauss-Bonnet term reduces
to topological invariants, rendering the theory dynamically equivalent to GR.
As a result, any attempt to probe the nontrivial effects of Lovelock terms
necessarily requires higher-dimensional spacetimes.
The emergence of extra dimensions should therefore not be viewed as an
independent assumption, but rather as an unavoidable consequence of incorporating
higher-curvature corrections into a consistent gravitational EFT.
To reconcile such higher-dimensional theories with observations, one is led
naturally to consider compactifications of the extra dimensions.

The exploration of spontaneous compactification in Lovelock gravity has a long history, dating back to early pioneering works such as \cite{MULLERHOISSEN1985106}, which already recognized that higher-curvature terms can play a crucial role in stabilizing extra dimensions. Since then, a substantial literature has developed, particularly on the cosmological dynamics of Lovelock theories. Much of this work has focused on time-dependent solutions and dynamical compactification scenarios, investigating whether cosmological evolution can drive the system toward a late-time attractor consisting of a stabilized internal space together with a four-dimensional expanding universe, often of de Sitter type. A comprehensive overview of this line of research and further references can be found in the recent review \cite{Pavluchenko:2024lcl}.

Despite the extensive body of work on Lovelock compactifications,
it has been increasingly recognized since the mid-2000s that
the consistency of the compactified vacuum cannot be taken for granted
\cite{Calcagni:2006ye,Charmousis:2008ce,Pavluchenko:2015daa}.
Although Lovelock gravity is often celebrated for yielding second-order
field equations, this property alone does not guarantee a ghost-free
spectrum in the four-dimensional effective theory obtained after
compactification.

This issue has been sharply highlighted in recent cosmological analyses.
In particular, Ref.~\cite{DeFelice2024} demonstrated that in dynamical
compactifications of Einstein--Gauss--Bonnet gravity, the tensor sector
generically suffers from ghost pathologies.
This result strongly suggests that the ghost problem in the Gauss–Bonnet sector is not merely an artifact of specific cosmological ansätze or time-dependent
solutions, but reflects a deeper obstruction in the compactified theory.

These developments suggest that judging the viability of a compactification
solely by the existence of cosmological attractors is insufficient.
From an effective-field-theory viewpoint, a more basic question must be
addressed first: whether the would-be vacuum itself is a consistent and
healthy background. Since the higher-dimensional Planck and Kaluza--Klein
scales are parametrically separated from the cosmological Hubble scale,
it is sufficient—and indeed essential—to analyze the Minkowski limit.
Only after establishing a ghost-free and locally stable Minkowski vacuum
can de Sitter backgrounds be regarded as controlled small-curvature
deformations around it.

In light of the tensor-sector pathologies of Einstein–Gauss–Bonnet compactifications, it is natural to ask whether the inclusion of the cubic Lovelock term can genuinely cure these instabilities. Indeed, several recent studies have reported apparently stable compactified solutions in cubic Lovelock models \cite{Chirkov:2018xrd,Pavluchenko:2024lcl}. However, a systematic and analytic assessment of the health and consistency of such vacua—especially at the level of the Minkowski limit and the tensor kinetic term—has so far been lacking.

In this paper, we undertake such an analysis. Adopting a particle-physics and EFT-oriented viewpoint, we establish, in a controlled and analytic manner, the conditions under which a ghost-free Minkowski vacuum exists in Lovelock gravity. We deliberately work in the Minkowski limit to isolate the essential structure of the theory, stripping away unnecessary cosmological complications while retaining all features relevant to stability and unitarity. At the same time, we go beyond the commonly assumed spherical internal spaces and consider compactification on the full class of compact irreducible Riemannian symmetric spaces (CIRS) \cite{Cartan1926,Cartan1927,Helgason1977}.

This generalization is motivated by two considerations. First, as we shall show, a ghost-free Minkowski vacuum is only possible when the internal dimension satisfies $\hat d \ge 6$, naturally forcing us beyond the simplest spherical examples. Second, CIRS spaces are homogeneous and highly symmetric, possessing no shape moduli other than an overall size modulus. As a result, the higher-dimensional dynamics consistently reduces to a single-field moduli problem. 
This setup allows us to isolate the genuinely geometric origin of stability, encoded in curvature invariants, while avoiding artifacts of overly restrictive assumptions.

Based on an exhaustive evaluation of Lovelock invariants up to cubic order for all CIRS classes, we uncover a universal mathematical property: the inverse Lovelock terms satisfy a log-convexity condition. This property leads to a universal no-go theorem for the Einstein--Gauss--Bonnet sector already at the level of the Minkowski vacuum,  while confirming that the cubic term can indeed allow for a locally ghost-free Minkowski vacuum within a narrow region of parameter space. Crucially, however, this systematic approach also allows us to identify a fundamental global issue that has been overlooked: even when a vacuum is locally stable, the global structure of the moduli space often 
renders it metastable and subject to what we term the "robust AdS preference."

Finally, we identify a mechanism by which this tendency can be evaded.
For an extremely restricted range of parameters, a kinetic barrier emerges that
pushes the unwanted AdS vacuum to the infinite boundary of the moduli space,
effectively isolating the Minkowski vacuum.
We show that this mechanism is exclusive to higher-rank CIRS spaces and is
fundamentally absent in rank-1 spaces such as spheres.
By identifying an infinite class of higher-rank internal geometries that realize
this scenario, we establish, for the first time, genuinely stable and ghost-free
Minkowski vacua in Lovelock gravity.

The remainder of this paper is organized as follows.
In Section~\ref{sec:Compactification}, we review the Lovelock gravity framework and the geometric
properties of CIRS spaces.
Section~\ref{sec:CIRS} is devoted to the explicit computation of Lovelock invariants and the
derivation of the universal log-convexity.
In Section~\ref{sec:MetastableVacua}, we establish the no-go theorem for the
Einstein--Gauss--Bonnet sector and analyze the local stability of Minkowski
vacua with cubic Lovelock terms.
Section~\ref{sec:GlobalStability} addresses the global structure of the moduli space and introduces the
kinetic barrier mechanism.
Section~\ref{sec:summary} contains our conclusions and discussion.
Supplementary technical results are collected in the Appendices. 
Appendix~\ref{sec:calcI} provides group-theoretic notations, and the systematic derivation of their values. 
In Appendix~\ref{sec:RobustAdS}, we extend our analysis to five-term Lovelock models and demonstrate the robust AdS preference specifically for \( S^n \) and \( \mathbb{C}P^m \) internal spaces.

\paragraph{Healthiness of the effective theory}

In this paper, we discuss the healthiness of the four-dimensional effective
theory in which the four-dimensional graviton is coupled to the size-modulus
scalar field arising from the compactification.
In particular, we require the absence of ghostlike excitations
in the propagating degrees of freedom, together with the stability of the
scalar sector.

Depending on the context, we sometimes distinguish between \emph{local} and
\emph{global} notions of healthiness.
Roughly speaking, the former refers to the absence of perturbative instabilities
around a given vacuum, while the latter concerns stability against large
excursions in the moduli space, including possible decay toward competing vacua.
The qualifiers \emph{local} and \emph{global} will be used accordingly when such
a distinction becomes relevant.

\section{Lovelock compactification}\label{sec:Compactification}
\subsection{Lovelock gravity}
\def\LL{{\cal R}}
\def\LG{{\cal G}}
To analyze compactifications in higher-curvature gravity, we begin by briefly reviewing the Lovelock theory of gravity.
We consider a $d$-dimensional spacetime manifold ${\cal M}$ equipped with a metric $g_{\mu\nu}$.
The Lovelock action is given by
\begin{align}
   S=\int_{\cal M} d^d x\, \sqrt{-g} 
   \sum_{k=0}^{\lfloor d/2\rfloor} \alpha_k \LL_k.
\end{align}
Throughout this paper, we assume that ${\cal M}$ is smooth and without boundary, 
so that total derivative terms do not contribute to the variation of the action.

Each $\LL_k$ is the $k$-th Lovelock term, defined by
\begin{align}
    \LL_k := \frac{1}{2^k}\,
    \delta^{\rho_1\sigma_1\cdots \rho_k\sigma_k}_{\mu_1\nu_1\cdots\mu_k\nu_k}
    \prod_{j=1}^k R^{\mu_j\nu_j}{}_{\rho_j\sigma_j},
\end{align}
where the generalized Kronecker delta\footnote{
They satisfy the useful identity, for example,
\begin{align}
     \delta^{\mu_1\mu_2\cdots \mu_n\lambda_1\lambda_2\cdots\lambda_m}
     _{\nu_1\nu_2\cdots\nu_n\lambda_1\lambda_2\cdots\lambda_m}
     =\frac{(d-n)!}{(d-m-n)!} 
     \delta^{\mu_1\mu_2\cdots \mu_n}_{\nu_1\nu_2\cdots\nu_n}.
\end{align}
} is given by
\begin{align}
    \delta^{\mu_1\mu_2\cdots \mu_p}_{\nu_1\nu_2\cdots\nu_p}
    := p! \,\delta^{\mu_1}_{[\nu_1}\delta^{\mu_2}_{\nu_2}\cdots 
       \delta^{\mu_p}_{\nu_p]}.
\end{align}
For instance, the zeroth, linear and quadratic Lovelock terms correspond to
the identity (cosmological term), the Ricci scalar, and the Gauss–Bonnet term, respectively:
\begin{align}
    \LL_0 = 1, \qquad
    \LL_1 = R, \qquad
    \LL_2 = R_{\mu\nu\rho\sigma}R^{\mu\nu\rho\sigma}
           -4R_{\mu\nu}R^{\mu\nu} + R^2.
\end{align}
Hence, $\alpha_1$ is related to the $d$-dimensional Einstein gravitational constant $\kappa_d$,
and $\alpha_0$ is proportional to the cosmological constant $\Lambda_d$:
\begin{align}
    \alpha_1 = \frac{1}{2\kappa_d} > 0, \qquad
    \alpha_0 = -\frac{\Lambda_d}{\kappa_d}.
\end{align}

The variation of the action yields
\begin{align}
    \delta S
    = -\int_{\cal M} d^d x \sqrt{-g}\,
      \LG^{\mu\nu} \delta g_{\mu\nu}, \qquad {\rm with~}\partial{\cal M}=0,
\end{align}
where
\begin{align}
    \LG^{\mu\nu}
    = \sum_{k=0}^{\lfloor d/2\rfloor} \alpha_k \LG_k{}^{\mu\nu}, \qquad
    \LG_k{}^\alpha{}_\beta
    := -\frac{1}{2^{k+1}}
       \delta^{\alpha\rho_1\sigma_1\cdots\rho_k\sigma_k}
              _{\beta\mu_1\nu_1\cdots\mu_k\nu_k}
       \prod_{j=1}^k R^{\mu_j\nu_j}{}_{\rho_j\sigma_j}
    = \LG_k{}_\beta{}^\alpha.
\end{align}
The tensor $\LG_k{}^{\mu\nu}$, hereafter referred to as the $k$-th order Lovelock tensor, is a generalized Einstein tensor,
which identically satisfies
\begin{align}
    \nabla_\nu \LG_k{}^{\mu\nu} \equiv 0,
    \qquad
    \LG_k{}^\lambda{}_\lambda \equiv -\frac{d-2k}{2}\,\LL_k.
\end{align}

By definition, both $\LL_k$ and $\LG_k^{\mu\nu}$ vanish identically in low dimensions:
\begin{align}
  \LL_k \equiv 0 \quad (d<2k), \qquad
  \LG_k^{\mu\nu} \equiv 0 \quad (d<2k+1).
\end{align}
Thus, for even dimensions $d=2m$, the highest-order Lovelock term is topological.
In particular, for $d=4$ the Gauss–Bonnet term\footnote{
We use the term \emph{Gauss--Bonnet term} to denote the quadratic Lovelock invariant $\LL_2$,
even though in the mathematical literature it often refers to the top Lovelock term in even dimensions.} becomes a topological invariant,
and the Lovelock theory reduces to Einstein gravity at the level of equations of motion.
To apply Lovelock gravity as an effective 4D theory,
we therefore consider compactification of the higher-dimensional manifold ${\cal M}$.

\subsection{Symmetric-space ansatz and irreducibility} \label{sec:irreduciblity}
Let us compactify the extra dimensions of the spacetime of the Lovelock theory, focusing on late-time attractor solutions corresponding to stationary de Sitter vacua.
We assume that the spacetime $\cal M$ forms a product of submanifolds as 
\begin{align}
    {\cal M}=\ul{\cal M}\times \prod_{i=1}^{p} \hat {\cal M}_{(i)}=\prod_{i=0}^{p} \hat {\cal M}_{(i)}, 
\end{align}
with their dimensions
\begin{align}
    d=\ul{d}+\sum_{i=1}^{p}\hat d_{(i)}=\sum_{i=0}^{p}\hat d_{(i)}, \quad 
    \hat d_{(i)}:= \dim \hat {\cal M}_{(i)},\quad \ul d=\hat d_{(0)}=4,
\end{align}
where each $\hat {\cal M}_{(i)}\,(i\not =0)$ is a compact Riemann manifold,
whereas $\ul{\cal M}=\hat {\cal M}_{(0)}$ is a non-compact pseudo-Riemann manifold as the spacetime of the universe.

Here, for simplicity, let us furthermore assume that $\cal M$ is homogeneous and its metric is given by
\begin{align}
    ds^2=g_{\mu\nu}dx^\mu dx^\nu=\sum_{i=0}^{p} \rho_{(i)}^2 d\hat s^2_{(i)},\qquad 
    d\hat s^2_{(i)}:=\hat g_{\mu_i\nu_i}^{(i)}(x_{(i)})dx^{\mu_i}_{(i)}dx^{\nu_i}_{(i)}
\end{align}
where each $\rho_{(i)}$ is a constant size modulus of ${\cal \hat M}_{(i)}$, whereas
$\rho_{(0)}$ is the inverse of the Hubble constant $H$.
Under this assumption, the $k$-th order Lovelock term is decomposed as
\begin{align}
 \LL_k(g)= k!\sum_{\{m_i\}\in \sigma_{k,p+1}} \prod_{i=0}^{p}
 \frac{\LL_{m_i}(\hat g_{(i)})}{m_i!\rho_{(i)}^{2m_i}},
\end{align}
where $\sigma_{k,p+1}$ is the set of weak compositions of $k$ into $p+1$ parts.
Here we have added arguments $g,\hat g_{(i)}$ to the Lovelock terms to clarify which metric it is defined using, and we will use this notation throughout this paper.
For consistency, the set $\{\rho_{(i)}\}$ must take values at saddle points of the following "potential":
\begin{align}
    V^{\rm tot}(\{\rho_{(i)}\}):=-\sum_{k=0}^{\lfloor d/2\rfloor} \alpha_k k!\sum_{\{m_i\}\in \sigma_{k,p+1}} \prod_{i=0}^{p}\rho_{(i)}^{\hat d_{(i)}-2m_i}\frac{\LL_{m_i}(\hat g_{(i)})}{m_i!},
\end{align}
which is a potential in the usual sense for size moduli $\rho_{(i)}$ 
other than $\rho_{(0)}=H^{-1}$, and therefore must yield at least a local minimum value for those.
Unlike the internal moduli, $\rho_{(0)}$ is not treated as a dynamical modulus.\footnote{The hubble parameter is a composite object as $H=\dot a/a$ and is not treated as other size moduli.}
The variation with respect to $H$ should be understood as imposing the constraint equation rather than a dynamical equation of motion.
Note that due to the assumption of spacetime homogeneity, it suffices to discuss only a single point in spacetime.

At this stage, the issue of vacuum stability—assuming a de Sitter external spacetime—is controlled  by the behavior of the moduli potential $V^{\rm tot}$ along individual modulus directions, independently of how many internal factors are present.
Before discussing products of internal manifolds, it is crucial to establish the stability of compactifications with a single internal factor ($p=1$), especially when the internal space has positive curvature. 
If stability cannot be achieved already in this minimal setting, introducing additional internal factors only increases the number of moduli and generically makes stabilization more difficult. 
Therefore, in what follows, we restrict ourselves to the case $p=1$ and focus on identifying internal geometries that admit stable vacua.

The saddle points of $V^{\text{tot}}$ mentioned above do not necessarily solve the full equations of motion for an arbitrary homogeneous manifold. However, if the internal manifold $({\cal \hat M},\hat g)$ is an irreducible symmetric space, its high degree of symmetry ensures that any rank-2 tensor invariant under the isometry group must be proportional to the metric. Specifically, the $k$-th order Lovelock tensor $\LG_k{}^\mu{}_{\nu}$ must satisfy
\begin{align}
    \LG_k{}^\mu{}_{\nu}(\hat g)=\frac{1}{\hat d}\delta^\mu_\nu\, 
    \LG_k{}^\lambda{}_{\lambda}(\hat g)
    =-\frac{\hat d-2k}{2\hat d}\delta^\mu_\nu\, \LL_k(\hat g),
    \label{eq:tensor_reduction}
\end{align}
where we have used the trace identity for the $k$-th Lovelock tensor and have omitted the index $(i)$. 
Consequently, an irreducible symmetric space is always an Einstein manifold.
Under this condition, the contribution to the equations of motion from each internal sector can be expressed as a derivative of the "potential" with respect to the size modulus:
\begin{align}
  \rho^{\hat d} \sum_k \tilde \alpha_k \LG_k{}^\mu{}_{\nu}(\rho^2\hat g)=- \frac{\delta^\mu_\nu}{2\hat d}\, \rho \frac{d}{d\rho} \left(\rho^{\hat d}\sum_k\tilde \alpha_k\LL_k(\rho^2\hat g)\right),
\end{align}
This identity implies that the full higher-dimensional equations of motion for the metric effectively reduce to the extremization conditions of the potential $V^{\text{tot}}$ with respect to the moduli ${\rho_{(i)}}$.
While maximally symmetric spaces (such as spheres $S^n$ or de Sitter spaces $dS_4$) satisfy Eq.~\eqref{eq:tensor_reduction} by definition, general homogeneous spaces do not necessarily possess this property. 

In this paper, we assume $p=1$ for simplicity and focus on cases where the internal space $\hat{\mathcal{M}}$ is a compact irreducible Riemannian symmetric space (CIRS). This class of spaces is broad enough to include $S^n$ and $\mathbb CP^m$ but restrictive enough to ensure that the dynamics of the extra dimensions are fully captured by a single-field moduli potential. As will be shown in Sec.~\ref{sec:GlobalStability}, the rich algebraic structure of CIRS spaces with higher rank provides the necessary freedom to achieve global stability, which is unattainable with simple spheres.

\section{Compact irreducible symmetric spaces (CIRS spaces)}\label{sec:CIRS}
\def\ad#1{{{\rm ad}(#1)}}
In this section, we examine compact irreducible Riemannian symmetric spaces (CIRS spaces),
which provide a natural generalization of spherical internal spaces,
as motivated in the previous section, to prepare for a thorough discussion of the Lovelock compactification.

\subsection{$G$-invariant metric and Lovelock invariants}
A compact Riemannian symmetric space is introduced as follows. 
The assumption of irreducibility will be required for calculations in later subsections, where explicit expressions of some Lovelock terms are established.

A Riemannian symmetric space can be realized as a homogeneous space 
\begin{align}
    \hat {\cal M}=G/H,\quad \hat d:=\dim \hat {\cal M}=\dim G-\dim H,
\end{align}
where $G$ is a Lie group acting transitively on the manifold and $H$ is the isotropy group at a chosen base point. Here $G/H$ denotes the space of left cosets, defined by the equivalence relation $g \sim gh$ for $h\in H$, on which $G$ acts transitively by left multiplication.
In the following, we restrict ourselves to a realization of the CIRS as a homogeneous space $G/H$ where $G$ is compact.
This can always be achieved since any compact Riemannian symmetric space admits a transitive action of a compact Lie group by isometries.
Therefore, let $T_{\hat A}$ be a basis of $\mathfrak g$ consisting of anti-Hermitian generators satisfying
\begin{align}
{} [T_{\hat A},T_{\hat B}]=f_{\hat A\hat B}{}^{\hat C}T_{\hat C}, \quad
\tr[T_{\hat A}T_{\hat B}]=-\delta_{\hat A\hat B},
\end{align}
with the structure constants $f_{\hat A\hat B}{}^{\hat C}$.

A defining property of symmetric spaces is the existence of an involutive automorphism $\sigma : G \to G$, satisfying $\sigma^2 = \mathrm{id}$. 
The subgroup $H$ is characterized as the set of elements invariant under $\sigma$.
At the level of the Lie algebra, the involution induces, so-called, a Cartan decomposition
\begin{align}
 \mathfrak g = \mathfrak h \oplus \mathfrak m ,   
\end{align}
where $\mathfrak h$ and $\mathfrak m$ are the eigenspaces of $\sigma$ with eigenvalues $+1$ and $-1$, respectively. 
Here $\mathfrak m$ carries a representation of $\mathfrak h$ via the adjoint action.
Corresponding to this decomposition, we split the algebra as 
$T_{\hat A}=T_A\oplus T_a$ where $T_A\in \mathfrak h, T_a\in \mathfrak m$ and thus they satisfy
\begin{align}
    [T_A,T_B]=f_{AB}{}^CT_C, \quad [T_A,T_a]=f_{Aa}{}^bT_b,\quad [T_a,T_b]=f_{ab}{}^AT_A.
\end{align}

A $G$-invariant metric can be constructed as follows.
Let us define the following one form, with $g\in G$,
\begin{align}
  \theta &:= \frac12(g^{-1}dg - \sigma(g^{-1}dg)),   
\end{align}
which is anti-Hermitian $\theta^\dagger=-\theta$ and 
is $\mathfrak m$-valued by construction, i.e. $\sigma(\theta)=-\theta$.
Then, a $G$-invariant metric and a $G$-covariant vielbein 
can be constructed as
\begin{align}
ds^2 &= \hat g_{\mu\nu} dx^\mu dx^\nu := -\tr[\theta^2] = \delta_{ab} e^a e^b, \quad
e^a := -\tr[T^a \theta].
\end{align}  
Under a left multiplication by any $g_0 \in G$,
a compensating right action by some $h\in H$ may be needed to restore
the coset representative. Since $\theta$ transforms covariantly under
this right $H$-action, the trace $\tr[\theta^2]$ remains unchanged,
so that the metric is invariant under the full action of $G$.

Choosing an involutive automorphism $\sigma$ naturally singles out the base point $o=[e]\in G/H$, with respect to which the Cartan decomposition is defined.
In a neighborhood of $o$, any element $g\in G$ can be uniquely written as
$g=e^Xe^\lambda \sim e^X$,
where $X\in\mathfrak m$ and $\lambda\in\mathfrak h$.
This provides local coordinates $x^\mu$ on $G/H$ through
$x^\mu = -\tr[X T_a]\delta^{a\, \mu}$.
Using this local coordinate system,  $\theta$ is rewritten as
\begin{align}
    \theta\sim\frac{\sinh \ad X}{\ad X}dX=dX+\frac16[X,[X,dX]]+\dots,
\end{align}
with $\ad X Y=[X,Y]$.
Here we used\footnote{
   The integral is taken with respect to $t$, whereas $dX$ denotes the exterior
derivative of $X$.}
\begin{align}
    \frac{1}{2}\left(e^{-X}de^{X}+de^Xe^{-X}\right)
    =\frac{1}{2}\int_0^1dt \left(e^{-t\, \ad X}+e^{t\,\ad X}\right)dX
    =\frac{\sinh \ad X}{\ad X}dX.
\end{align}
Therefore,  we can calculate the Riemann curvature tensor at the origin $o$
and the homogeneity of $G/H$ guarantees
that the result at $o$ can be extended to be everywhere through 
completion by the vielbein $e^a=e^a{}_\mu dx^\mu$ as
\begin{align}
    R_{\mu\nu\rho\sigma}(\hat g)
    =e^a{}_{\mu}e^b{}_{\nu}e^c{}_\rho e^d{}_\sigma \times f_{abA}f_{cd}{}^A. \label{eq:curvature}
\end{align}
Here we lower the indices by $\hat g_{\mu\nu},\delta_{ab},\delta_{AB}$ and raise them by their inverse. 
Since $\hat{\cal M}=G/H$ is assumed to be a compact Riemannian symmetric space,
the metric $\hat g_{\mu\nu}$ constructed above is positive definite.
Consequently, the induced inner products on $\mathfrak m$ and $\mathfrak h$,
given by $\delta_{ab}$ and $\delta_{AB}$, are also positive definite.
This positivity plays a crucial role in the following.
In particular, owing to the positive definiteness of the metric,
the Lovelock terms $\LL_k$ on $\hat{\cal M}$ turn out to be positive semidefinite.
This can be seen by rewriting them as
\begin{align}
    \LL_k(\hat g)=\Psi_{a_1a_2\cdots a_{2k}}{}^{A_1A_2\cdots A_k}
    \Psi^{a_1a_2\cdots a_{2k}}{}_{A_1A_2\cdots A_k}\ge 0
\end{align}
with
\begin{align}
   \Psi_{a_1a_2\cdots a_{2k}}{}^{A_1A_2\cdots A_k}:= 
   \frac1{\sqrt{2^k(2k)!}}\delta_{a_1a_2\cdots a_{2k}}^{b_1c_1\cdots b_kc_k}\prod_{j=1}^k f_{b_jc_j}{}^{A_j}.
\end{align}

Note that formula \eqref{eq:curvature} has a constant-multiple indeterminacy due to rescaling the metric or changing the representation of $G$. 
Different choices of the representation of $G$ only affect the overall
normalization of the invariant trace and hence rescale the metric by a constant.
This motivates our focus on scale-invariant quantities below.
The rescaling of the metric causes the rescaling of the curvatures and, in particular, the Lovelock terms and the volume change as
\begin{align}
    \LL_k(\rho^2\hat g)=\rho^{-2k} \LL_k(\hat g), \quad {\rm Vol}(\hat {\cal M},\rho^2\hat g)=
    \rho^{\hat d}{\rm Vol}(\hat {\cal M},\hat g)
\end{align}
and thus, it is convenient to consider the following scale invariant quantities:
\begin{align}
    \gamma_k(\hat {\cal M}):= \frac{\LL_k(\hat g) R(\hat g)}{\LL_{k+1}(\hat g)},
    \quad {\rm Vol}(\hat {\cal M},\hat g)(R(\hat g))^{\frac{\hat d}{2}}.
\end{align}
For $\hat d=2m$, the Euler characteristic $\chi(\hat {\cal M})$ is calculated by integrating the highest Lovelock term as
\begin{align}
    \chi(\hat {\cal M})=\frac{1}{(4\pi)^m m!}\int_{\hat {\cal M}}d^{2m}x\sqrt{\hat g}\LL_m(\hat g)=\frac{{\rm Vol}(\hat {\cal M},\hat g)\,\LL_m(\hat g)}{(4\pi)^m m!}\ge 0.
\end{align}
\subsection{Examples with $S^n$ and $\mathbb CP^m$}

As simple examples where all Lovelock terms $\LL_k$ can be immediately computed, we consider the $n$-dimensional unit sphere $S^n$ and the $m$-dimensional complex projective space $\mathbb CP^m$. 
Let us temporarily fix the normalization conventions of 
their well-known metrics:
\begin{align}
    ds^2(S^n)& = |d\vec v|^2 \quad {\rm with~} \vec v \in \mathbb R^{n+1}, \quad |\vec v|^2=1, \\
    ds^2(\mathbb CP^m)&= 2\hat g_{\alpha\bar\beta }dz^\alpha d\bar z^{\bar \beta}=2\frac{\p^2\log(1+|\vec z|^2)}{\p z^\alpha\p \bar z^{\bar \beta}} dz^\alpha d\bar z^{\bar \beta}, \quad {\rm with~} \vec z \in \mathbb C^m.
\end{align}
Using these normalizations, the corresponding Riemann curvature tensors take the simple forms
\begin{align}
    R^{\mu\nu}{}_{\rho\sigma}(S^n)=\delta^\mu_\rho \delta^\nu_\sigma-\delta^\mu_\sigma\delta^\nu_\rho
,\quad R_{\alpha \bar \beta \bar \gamma \delta}(\mathbb CP^m)=\hat g_{\alpha\bar \gamma }\hat g_{\bar \beta\delta}+\hat g_{\alpha\bar \beta}\hat g_{\bar \gamma \delta},
\end{align}
which are constant and compatible with the symmetric space structure, allowing straightforward calculation of Lovelock terms.

For these spaces, the Lovelock densities $\LL_k$ can be computed explicitly:
\begin{align}
    \LL_k(S^n)=\frac{n!}{(n-2k)!},\quad \LL_k(\mathbb CP^m)=\frac{2^km!(m+1)!}{(m-k)!(m-k+1)!}.\label{eq:SCPLL}
\end{align}
These lead to the scale-invariant ratios
\begin{align}
    \gamma_k(S^n)=\frac{n(n-1)}{(n-2k)(n-2k-1)}\ge 1,\quad \gamma_k(\mathbb CP^m)=\frac{m(m+1)}{(m-k)(m-k+1)}\ge 1.\label{eq:gammaSnCPm}
\end{align}
For the chosen normalization, 
the corresponding volumes are
\begin{align}
    {\rm Vol}(S^n)=\frac{2\pi^{\frac{n+1}{2}}}{\Gamma\left(\frac{n+1}{2}\right)},
    \quad {\rm Vol}(\mathbb CP^m)=\frac{(2\pi)^m}{m!}
\end{align}
from which we can also verify their Euler characteristics: $\chi(S^{2m})=2,\chi(\mathbb CP^m)=m+1$.

These examples are chosen for their simplicity: all $\LL_k$ can be explicitly calculated. 
For general symmetric spaces, we will present formulas for $k=1,2,3$ in the next section; compact expressions for arbitrary $k$ are not generally available, or at least not known to the author.
\subsection{Explicit formulas for $\LL_1,\LL_2$ and $\LL_3$}
CIRS spaces have been classified by \'Elie Cartan \cite{Cartan1926,Cartan1927,Helgason1977}.
In modern approaches, they are broadly categorized into two classes: class A and class B.

\subsubsection{Class A: simple $G$}

To calculate $\LL_k$, let us begin with the case of class A, where the group $G$ of the coset $G/H$ is simple. All CIRS spaces in class A are listed in Table \ref{tab:Ihvaluelist},
where rank-1 spaces $S^n$ and $\mathbb CP^m$ belong to BDI and AIII, respectively.
The structure constants are normalized as
\begin{align}
    \frac12 f^{\hat A\hat C\hat D}f_{\hat B\hat C\hat D}
    =N_{\mathfrak g}({\rm ad})\,\delta^{\hat A}_{\hat B},
    \label{eq:Normal-g}
\end{align}
where $N_{\mathfrak g}({\rm ad})>0$ is a normalization constant.
Using this normalization, Formula \eqref{eq:curvature} shows that a CIRS space (in class A) is an Einstein manifold:
\begin{align}
    R_{\mu \nu}(\hat g)
    =f_{acA}f_{b}{}^{cA}e^a{}_{\mu}e^b{}_\nu
    =\frac12f_{a\hat A\hat B}f_{b}{}^{\hat A\hat B}e^a{}_{\mu}e^b{}_\nu
    =N_\mathfrak g({\rm ad})\, \hat g_{\mu\nu}.
\end{align}
Hence,
\begin{align}
    \LL_1(\hat g)=R(\hat g)=N_{\mathfrak g}({\rm ad})\,\hat d,
    \quad {\rm with}\ \hat d={\rm dim}\,\mathfrak m.
\end{align}
Consequently, the curvature invariants containing the Ricci tensor are computed as
\begin{align}
    &R^{\mu\nu}(\hat g)R_{\mu\nu}(\hat g)
      =[N_{\mathfrak g}({\rm ad})]^2\hat d, \quad 
      R^\mu{}_\nu (\hat g)R^\nu{}_\rho (\hat g)R^\rho{}_\mu(\hat g)
      =[N_{\mathfrak g}({\rm ad})]^3\hat d, \nonumber \\
   &R^{\mu\nu}{}_{\rho\sigma}(\hat g)R^\rho{}_{\mu}(\hat g)R^\sigma{}_\nu(\hat g)
      =[N_{\mathfrak g}({\rm ad})]^3\hat d.
\end{align}

For the explicit evaluation of $\LL_2$ and $\LL_3$, it is necessary to decompose $\mathfrak h$ to obtain further information.
The isotropy algebra $\mathfrak h$ can always be written as the direct sum of simple and abelian parts:
\begin{align}
 \mathfrak{h} = \mathfrak{z} \oplus \bigoplus_{i=1}^{r} \mathfrak{h}_i
 =\bigoplus_{i=0}^{r} \mathfrak{h}_i,
\end{align}
where each $\mathfrak{h}_i$ ($i\neq0$) is simple and $\mathfrak{z}=\mathfrak h_0$ is the (possibly trivial) center. 
We introduce indices $A_{(i)}$ for $\mathfrak h_i$ so that
\begin{align}
    f_{abA}f_{cd}{}^A
    =(\tau_A)_{ab}(\tau^A)_{cd}
    =\sum_{i=0}^r(\tau_{A_{(i)}})_{ab}(\tau^{A_{(i)}})_{cd},
\end{align}
where $(\tau_A)_{ab}:=f_{abA}$ denotes the real antisymmetric generators of $\mathfrak h$ acting on $\mathfrak m$, satisfying $[\tau_A,\tau_B]=f_{AB}{}^C\tau_C$.
In what follows, we do not use the Einstein summation convention over the index $i$; summations are written explicitly.

For an irreducible symmetric space, in the Cartan decomposition $\mathfrak{g} = \mathfrak{h} \oplus \mathfrak{m}$, each simple component $\mathfrak{h}_i$ acts on $\mathfrak{m}$ as a real irreducible representation.\footnote{In some cases, as a complex representation, $\mathfrak{m}$ may decompose into a representation and its complex conjugate, but as a real representation, it remains irreducible.} Therefore, by Schur’s lemma for real representations, any $\mathfrak{h}_i$-invariant bilinear form on $\mathfrak{m}$ must be proportional to the metric:
\begin{align}
    f^{acA_{(i)}}f_{b c A_{(i)}}
    = -(\tau^{A_{(i)}}\tau_{A_{(i)}})^a{}_b
    = C_{\mathfrak h_i}(\mathfrak m)\,\delta^a_{b},
    \label{eq:Casimir}
\end{align}
where $C_{\mathfrak h_i}(\mathfrak m) > 0$ is the quadratic Casimir of $\mathfrak{h}_i$ in the representation on $\mathfrak{m}$.
Since $\mathfrak h_i$ is simple, its generators $(\tau_{A_{(i)}})_{ab}$ are normalized as
\begin{align}
    \frac{1}{2}f^{A_{(i)}ab}f_{B_{(j)}ab}
    =-\frac{1}{2}\tr[\tau^{A_{(i)}}\tau_{B_{(j)}}]
    =N_{\mathfrak h_i}(\mathfrak m)\,\delta^i_j
     \delta^{A_{(i)}}_{B_{(i)}}, 
     \label{eq:Normal-m}
\end{align}
and their adjoint representation is normalized as
\begin{align}
   \frac{1}{2}f^{A_{(i)}CD}f_{B_{(j)}CD}
   =-\frac{1}{2}\tr[\ad{\tau^{A_{(i)}}}\ad{\tau_{B_{(j)}}}]
   =N_{\mathfrak h_i}({\rm ad})\,\delta^i_j 
    \delta^{A_{(i)}}_{B_{(i)}},
    \label{eq:Normal-ad}
\end{align}
where $N_{\mathfrak h_i}(\mathfrak m),\,N_{\mathfrak h_i}({\rm ad})>0$, and their ratio
\begin{align}
    I_i=I_{\mathfrak h_i}(\mathfrak m)
    :=\frac{N_{\mathfrak h_i}(\mathfrak m)}{N_{\mathfrak h_i}({\rm ad})}
\end{align}
is independent of normalization. 
All values of $I_i$ for the class-A symmetric spaces are listed in Table~\ref{tab:Ihvaluelist},
and details of their computation can be found in Appendix \ref{sec:calcI}.
\begin{table}[t]
    \resizebox{\textwidth}{!}{
    \begin{tabular}{c||c|c|c|c|c|c|}
       &$\mathfrak g$ & $\mathfrak h$ &$\mathfrak m$ &$\hat d={\rm dim}\, \mathfrak m$ & $(I_i,\,{\rm dim }\mathfrak h_i)$&rank\\ \hline \hline 
      AI   & $\mathfrak{su}(n)$ &$\mathfrak{so}(n)$ & ${\rm Sym}^2_0({\bf n})$
      & $\frac{(n-1)(n+2)}2$ & $\left(\frac{n+2}{n-2},\frac{n(n-1)}2\right)$&$n-1$ \\ \hline 
       AII  & $\mathfrak{su}(2n)$& $\mathfrak{sp}(n)$ &$\Lambda^2_0({\bf 2n})$ &{\footnotesize$(n-1)(2n+1)$} & $\left(\frac{n-1}{n+1},n(2n+1)\right)$&$n-1$\\ \hline 
       AIII& $\mathfrak {su}(p+q)$& $\mathfrak z\oplus \mathfrak{su}(p)\oplus \mathfrak {su}(q)$
       & $({\bf p},{\bf \bar q})_+\oplus$ c.c. &$2pq$ & \footnotesize $(\infty,1),\left(\frac{q}{p},p^2-1\right), \left(\frac{p}{q}, q^2-1\right)$&min$(p,q)$\\ \hline 
       BDI&$\mathfrak {so}(p+q)$& $\mathfrak{so}(p)\oplus \mathfrak {so}(q)$& $({\bf p},{\bf q})$
       & $pq$ & $\left(\frac{q}{p-2},\frac{p(p-1)}2\right), \left(\frac{p}{q-2}, \frac{q(q-1)}{2}\right)$&min$(p,q)$\\ \hline 
       DIII&$\mathfrak{so}(2n)$& $\mathfrak z \oplus \mathfrak{su}(n)$ & 
       $\Lambda^2({\bf n})_+\oplus$ c.c. & $n(n-1)$& $(\infty,1), \,\left(\frac{n-2}{n}, n^2-1\right) $&$[n/2]$\\ \hline 
       CI&$\mathfrak{sp}(n)$& $\mathfrak z \oplus \mathfrak{su}(n)$ & ${\rm Sym}^2({\bf n})_+\oplus$ c.c.& $n(n+1)$ &
       $(\infty,1), \,\left(\frac{n+2}{n}, n^2-1\right) $&$n$\\ \hline 
       CII&$\mathfrak {sp}(p+q)$& $\mathfrak{sp}(p)\oplus \mathfrak {sp}(q)$ &$({\bf 2p},{\bf 2q})$
       & $4pq$ & \footnotesize $\left(\frac{q}{p+1},p(2p+1)\right), \left(\frac{p}{q+1}, q(2q+1)\right)$&min$(p,q)$\\ \hline 
       EI& $\mathfrak e_6$ & $\mathfrak {sp}(4)$ & ${\Lambda^4_0}({\bf 8}) $&$42$ & $\left(\frac75, 36\right)$&6\\ \hline 
       EII& $\mathfrak e_6$ & $\mathfrak {su}(6)\oplus \mathfrak{su}(2)$& 
       $\left(\Lambda^3(\bf 6),\bf 2\right)$& $40$&$(1,35),(5,3)$&4\\ \hline 
       EIII&$\mathfrak e_6$ & $\mathfrak z\oplus \mathfrak{so}(10)$ & ${\bf spin}_+\oplus$ c.c. &$32$&$(\infty,1),\left(\frac12,45\right)$&2\\ \hline 
       EIV&$\mathfrak e_6$ & $\mathfrak{f}_4$ & {\bf 26} &$26$&$\left(\frac13,52\right)$&2\\ \hline 
       EV&$\mathfrak e_7$ & $\mathfrak{su}(8)$ &${\Lambda^4}({\bf 8})$ &$70$&$\left(\frac54,63\right)$&7\\ \hline 
       EVI&$\mathfrak e_7$ & $\mathfrak{so}(12)\oplus \mathfrak{su}(2)$ & $({\bf spin},{\bf 2})$&$64$&$\left(\frac45,66\right),\,(8,3)$&4\\ \hline 
       EVII&$\mathfrak e_7$ & $\mathfrak z\oplus \mathfrak{e}_6$ & $\bf 27_+\oplus$ c.c.&$54$&$(\infty,1),\,\left(\frac12,78\right)$&3\\ \hline 
       EVIII&$\mathfrak e_8$ & $\mathfrak{so}(16)$ & ${\bf spin}$&$128$&$\left(\frac87,120\right)$&8\\ \hline 
       EIX&$\mathfrak e_8$ & $\mathfrak e_7\oplus \mathfrak{su}(2)$ & 
       $({\bf 56},{\bf 2})$&$112$&$\left(\frac23,133\right),\,(14,3)$&4\\ \hline 
       FI&$\mathfrak f_4$ & $\mathfrak{sp}(3) \oplus \mathfrak{su}(2)$ & $(\Lambda^3_0({\bf 6}),{\bf  2})$&$28$&$\left(\frac54,21\right),\,\left(\frac72,3\right)$&4\\ \hline 
       FII&$\mathfrak f_4$ & $ \mathfrak{so}(9)$ & {\bf spin}&$16$&$\left(\frac27,36\right)$&1\\ \hline 
       G&$\mathfrak g_2$ & $\mathfrak{su}(2)\oplus \mathfrak{su}(2)$ & $({\bf 4},{\bf 2})$&$8$&$(5,3),\,(1,3)$&2\\
       \hline 
    \end{tabular}
    }
    \caption{The $\mathfrak h$-representations $\mathfrak m$ and their index ratios $I_i$ for all class-A symmetric spaces.
In this table, $\mathfrak m$ is regarded as an $\mathfrak h$-module defined by the Cartan decomposition $\mathfrak g=\mathfrak h\oplus\mathfrak m$.
Representation notations (e.g., ${\rm Sym}^2_0({\bf n}), ({\bf p},{\bf \bar q})_+\oplus {\rm c.c.})$ are explained in Appendix \ref{sec:calcI}.
}
    \label{tab:Ihvaluelist}
\end{table}
Considering $\tr[\tau^{A_{(i)}}\tau_{A_{(i)}}]$ in two different ways using \eqref{eq:Casimir} and \eqref{eq:Normal-m}, we obtain the relation 
\begin{align}
    C_{\mathfrak h_i}(\mathfrak m)=\frac{2\,{\rm dim}\,\mathfrak h_i}{{\rm dim}\,\mathfrak m} 
    N_{\mathfrak h_i}(\mathfrak m).
\end{align}
Using Eq.\,\eqref{eq:Normal-m}, the square of the Riemann curvature tensor is evaluated as
\begin{align}
    R^{\mu\nu\rho\sigma}(\hat g)R_{\mu\nu\rho\sigma}(\hat g)
    =\sum_{i=0}^r\tr[\tau^{A_{(i)}}\tau^{B_{(i)}}]\,
     \tr[\tau_{A_{(i)}}\tau_{B_{(i)}}]
    =\sum_{i=0}^r [2N_{\mathfrak h_i}(\mathfrak m)]^2\,{\rm dim}\,\mathfrak h_i.
\end{align}
Similarly, by employing Eq.\,\eqref{eq:Normal-ad}, we find
\begin{align}
    [\tau_A,\,\tau_B]\otimes[\tau^A,\,\tau^B]
    =2\sum_{i=0}^r N_{\mathfrak h_i}({\rm ad})\,\tau_{A_{(i)}}\otimes \tau^{A_{(i)}}.
    \label{eq:tautautautau}
\end{align}

By decomposing the normalization in \eqref{eq:Normal-g} and comparing it with 
\eqref{eq:Casimir}, \eqref{eq:Normal-m}, and \eqref{eq:Normal-ad}, 
we obtain the following relations:
\begin{align}
    N_{\mathfrak g}({\rm ad})
    =\sum_{i=0}^r C_{\mathfrak h_i}(\mathfrak m),
    \qquad 
    N_{\mathfrak g}({\rm ad})
    =N_{\mathfrak h_i}({\rm ad})+N_{\mathfrak h_i}(\mathfrak m)
    \quad ({\rm for~each~}i).
\end{align}
From these two relations, we can express the normalization constants as
\begin{align}
    N_{\mathfrak h_i}({\rm ad})
    =\frac{N_{\mathfrak g}({\rm ad})}{1+I_{\mathfrak h_i}(\mathfrak m)},\qquad
    N_{\mathfrak h_i}(\mathfrak m)
    =\frac{I_{\mathfrak h_i}(\mathfrak m)\,N_{\mathfrak g}({\rm ad})}
           {1+I_{\mathfrak h_i}(\mathfrak m)},
\end{align}
and also find a consistency condition for the symmetric space $G/H$,
\begin{align}
    \sum_{i=0}^r
    \frac{2I_{\mathfrak h_i}(\mathfrak m)\,{\rm dim}\,\mathfrak h_i}
         {1+I_{\mathfrak h_i}(\mathfrak m)}
    ={\rm dim}\,\mathfrak m
    =\hat d.
    \label{eq:cosistency}
\end{align}
For the center, if it is non-trivial, $N_{\mathfrak h_0}({\rm ad})$ vanishes 
while $N_{\mathfrak h_0}(\mathfrak m)$ remains finite; 
we therefore formally set $I_0(\mathfrak m)=\infty$ with ${\rm dim}\,\mathfrak h_0=1$.
The same treatment applies to cases where $n=2$ in the AI series, or $p,q=2$ in the BDI series in Table \ref{tab:Ihvaluelist}. Readers can check that all series satisfy the above consistency condition by applying the data from Table \ref{tab:Ihvaluelist} while paying attention to these points.
There is a useful identity
\begin{align}
    \tau_B \tau_{A_{(i)}} \tau^B
    = -N_{\mathfrak h_i}(\mathfrak m)\,\tau_{A_{(i)}},
    \label{eq:tautautau}
\end{align}
which can be derived by combining the above consistency condition with the relations obtained so far. 
However, the simplest derivation proceeds as follows. 
Through Formula \eqref{eq:curvature}, 
the Bianchi identity $R_{a[bcd]}=0$ is realized as a part of the Jacobi identities for the structure constants $f_{\hat A\hat B\hat C}$:
\begin{align}
  0
  =(\tau_B)_{ab} (\tau^B)_{cd}
   +(\tau_B)_{ac} (\tau^B)_{db}
   +(\tau_B)_{ad} (\tau^B)_{bc}.
  \label{eq:Bianchi}
\end{align}
This relation corresponds to the property $[\mathfrak m,\mathfrak m]\subset\mathfrak h$ 
and guarantees that $\mathfrak h\oplus\mathfrak m$ forms a Cartan decomposition. 
Multiplying both sides of Eq.\,\eqref{eq:Bianchi} by $(\tau_{A_{(i)}})^{cd}$ 
immediately yields the identity \eqref{eq:tautautau}.

Substituting Formula \eqref{eq:curvature} into $\LL_2$, that is, the Gauss–Bonnet term, 
and using the relations discussed above, we finally obtain the formula
\begin{align}
    \frac{\LL_2(\hat g)}{R^2(\hat g)}
    &= 1-\frac{4}{\hat d}
       +\frac{4}{\hat d^2}
        \sum_{i=0}^r
        \frac{I_i^2\,{\rm dim}\,\mathfrak h_i}{(1+I_i)^2}\\
    &= 1-\frac{2}{\hat d}
       -\frac{4}{\hat d^2}
        \sum_{i=1}^r
        \frac{I_i\,{\rm dim}\,\mathfrak h_i}{(1+I_i)^2},\label{eq:lovelockR2}
\end{align}
where the transition from the first line to the second line 
uses the consistency condition \eqref{eq:cosistency}.
The left-hand side of the above equation is a scale-invariant quantity, 
and consequently the right-hand side is also independent of the normalization constants.
\begin{figure}[h]           
    \centering              
    \includegraphics[width=1.0\textwidth,trim=50 0 50 0, clip]{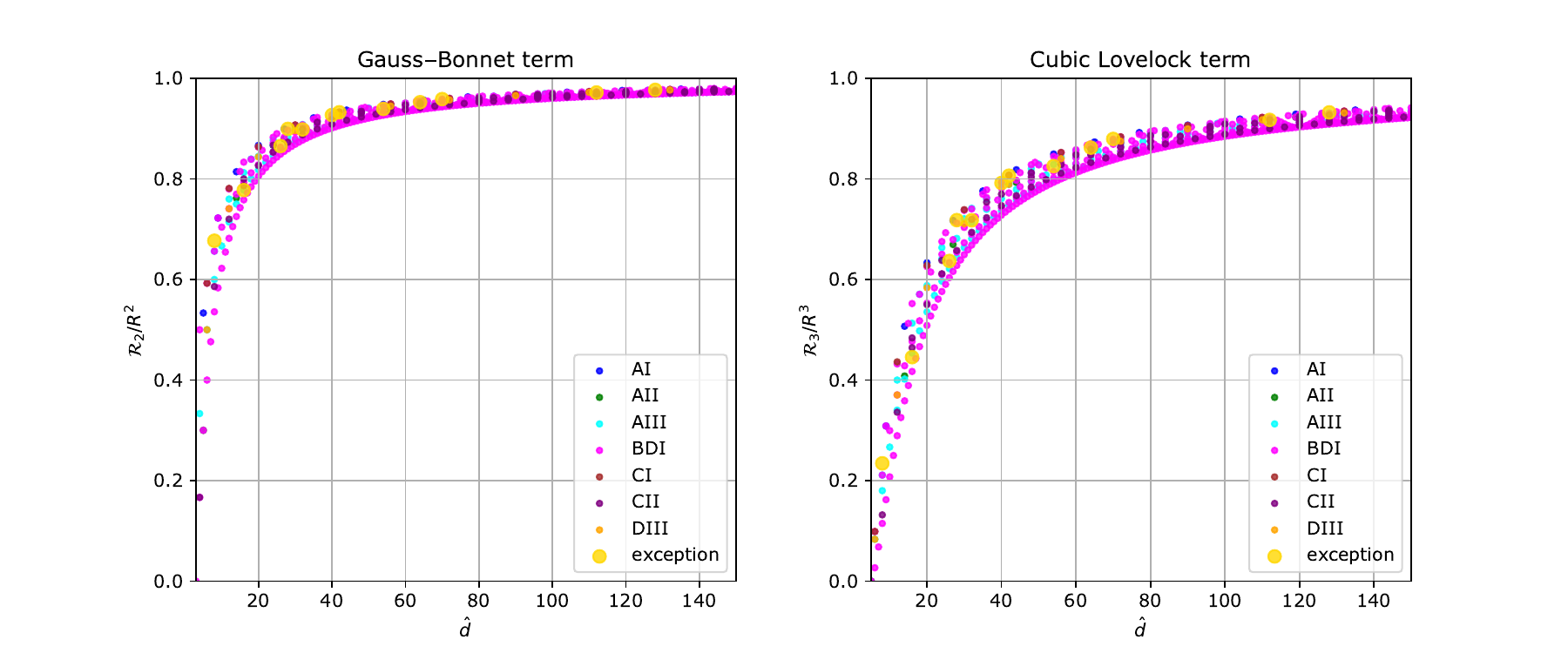} 
    \caption{$\hat d$-dependence of the Gauss-Bonnet term (left) and the cubic Lovelock term (right).} 
    \label{fig:Lovelockterms}      
\end{figure}
By applying the data from Table \ref{tab:Ihvaluelist} to the above formula, 
we calculated $\LL_2/R^2$ for all series of CIRS spaces and plotted all data points $(\hat d, \LL_2/R^2)$ for CIRS spaces with $\hat d \le 150$ in the left panel of Fig.\ref{fig:Lovelockterms}.

The cubic Lovelock term $\LL_3$ can be expressed explicitly 
in terms of the Riemann curvature tensor as
\begin{align}
    \LL_3
    &= 8R^{\mu\nu}{}_{\lambda\sigma}R_{\nu\rho}{}^{\tau\lambda}R^{\rho\sigma}{}_{\mu\tau}
     +2R^{\mu\nu}{}_{\rho\sigma}R^{\rho\sigma}{}_{\lambda\tau}R^{\lambda\tau}{}_{\mu\nu}
     -24 R^{\mu\nu}{}_{\rho\sigma}R^{\rho\sigma}{}_{\lambda\nu}R^{\lambda}{}_{\mu}\nonumber\\
    &\quad
     +24 R^{\mu\nu}{}_{\rho\sigma}R^\rho{}_{\mu}R^\sigma{}_\nu
     +16 R^\mu{}_\nu R^\nu{}_\rho R^\rho{}_\mu
     +3R\,\LL_2 - 2R^3.
\end{align}
Substituting Formula \eqref{eq:curvature} into the above expression, 
we can directly evaluate all terms on the right-hand side 
except for the first one:
\begin{align}
    R^{\mu\nu}{}_{\lambda\sigma}(\hat g)
    R_{\nu\rho}{}^{\tau\lambda}(\hat g)
    R^{\rho\sigma}{}_{\mu\tau}(\hat g)
    = -\tr[\tau_A\tau_B\tau_C\tau^A\tau^B\tau^C].
\end{align}
The remaining trace can be computed by employing 
Eqs.\,\eqref{eq:tautautautau} and \eqref{eq:tautautau}:
\begin{align}
    \tr[\tau_A\tau_B\tau_C\tau^A\tau^B\tau^C]
    &= \tr[\tau^B\tau_A\tau_B\tau_C\tau^A\tau^C]
     +\frac{1}{2}\tr[\tau_A[\tau_B,\,\tau_C]\tau^A[\tau^B,\,\tau^C]]\\
    &= -2\sum_{i=0}^r
       [N_{\mathfrak h_i}(\mathfrak m)]^2
       \left(N_{\mathfrak h_i}(\mathfrak m)
             -N_{\mathfrak h_i}({\rm ad})\right)
       {\rm dim}\,\mathfrak h_i.
\end{align}
Collecting all the results above, 
we arrive at the following compact expression:
\begin{align}
    \frac{\LL_3(\hat g)}{R^3(\hat g)}
    &= 1-\frac{6}{\hat d}
       +\frac{8}{\hat d^2}
       -2\!\left(\frac{6}{\hat d^2}
       -\frac{8}{\hat d^3}\right)
       \sum_{i=1}^r
       \frac{I_i\,{\rm dim}\,\mathfrak h_i}{(1+I_i)^2}
       +\frac{48}{\hat d^3}
        \sum_{i=1}^r
        \frac{I_i\,{\rm dim}\,\mathfrak h_i}{(1+I_i)^3}.\label{eq:LovelockR3}
\end{align}
In the right panel of Fig.\ref{fig:Lovelockterms},
we plotted all data points $(\hat d, \LL_3/R^3)$ 
for CIRS spaces with $\hat d \le 150$.

Note that  calculating the quartic or higher-order Lovelock term $\LL_k(k\ge 4)$
requires computing the squares of the high-order Casimir invariant tensors,
which are beyond the scope of this paper and will be addressed in future work.

\subsubsection{Class B: the compact simple Lie group}
Next, let us consider the group-type (class B) cases, i.e.\ the compact simple
Lie groups equipped with the bi-invariant metric
\begin{align}
    ds^2=-\tr[(g^{-1}dg)^2]=\delta_{AB}e^Ae^B,
    \qquad 
    e^A=-\tr[T^A g^{-1}dg], 
    \qquad g\in G.
\end{align}
 
A compact Lie group G can be viewed as a Riemannian symmetric space in two equivalent ways.
One is the standard representation $(G\times G)/\mathrm{diag}\,G$, where the involution acts as $(g_1,g_2)\mapsto(g_2,g_1)$.
Under the canonical identification $[(g_1,g_2)]\mapsto g_1g_2^{-1}$, this becomes the familiar involution on $G$ itself, $g\mapsto g^{-1}$, whose fixed point determines the origin of the expansion $g=e^X$ with $X\in\mathfrak{g}$.

The corresponding Riemann curvature tensor takes the form
\begin{align}
    R_{\mu\nu\rho\sigma}(\hat g)
    = e^A{}_{\mu} e^B{}_{\nu} e^C{}_{\rho} e^D{}_{\sigma}
      \, f_{ABE} f_{CD}{}^{E}. 
    \label{eq:curvatureB}
\end{align}
Comparing this with the class-A expression shows that the required
replacement is simply
\((\tau^A)^a{}_{b}=f^{Aa}{}_{b} \rightarrow
(\mathrm{ad}\,\tau^A)^B{}_{C}=f^{AB}{}_{C}\).
Thus one can reuse the class-A formulas by setting
\(r=1\), \(\dim\mathfrak{h}_i=\hat d\), and \(I_i=1\).
In particular, we immediately obtain the closed forms
\begin{align}
    \frac{\LL_2(\hat g)}{R^2(\hat g)} = 1-\frac{3}{\hat d},
    \qquad
    \frac{\LL_3(\hat g)}{R^3(\hat g)}
    =\left(1-\frac{3}{\hat d}\right)\left(1-\frac{6}{\hat d}\right).
\end{align}

Since the dimensions of compact simple Lie groups begin as
\(3,\,(6),\,8,\,10,\dots\),
one readily checks that these expressions vanish in the lowest-dimensional
cases, in accordance with the intrinsic dimensional constraints of the
Lovelock densities~\(\LL_k\).

The factor \(1-6/\hat d\) appearing in \(\LL_3\) admits a simple
interpretation.
It is well known that the Euler characteristic of a compact simple
Lie group vanishes, \(\chi(G)=0\).
Hence, for an even-dimensional group manifold \(\hat d=2m\), the top
Lovelock density satisfies \(\LL_m(\hat g)=0\).
Although \(SO(4)\) is not simple, formally applying the above expressions
still yields \(\LL_3(\hat g)=0\), consistent with the fact that
\(\chi(SO(4))=0\).

\subsection{Universal log-convexity on sequence $\gamma_k$}\label{sec:log-convexity}
Here, let us extract the key properties from the results obtained thus far.
For each CIRS space $\cal \hat M$, there exists 
the following sequence of normalization-independent quantities
consisting of the Lovelock invariants:
\begin{align}
    \gamma_k=\gamma_k({\cal \hat M}):= \frac{\LL_k(\hat g) R(\hat g)}{\LL_{k+1}(\hat g)},\quad 0\le 2k \le \hat d-2.
\end{align}
In later sections, not only the individual values of these quantities but also the relationships between them will become important.

For the sequences $\gamma_k$ associated with $S^n$ and $\mathbb{C}P^m$
given in \eqref{eq:gammaSnCPm},
one can show that the sequence $\gamma_k$
is completely log-convex, in the sense that all finite differences $\Delta^\ell \log \gamma_k$ are positive.\footnote{
For both of $S^n$ and $\mathbb{C}P^m$ the sequence $\gamma_k$ admits a  smooth real extension $\gamma(x)$ such that $f(x)=\log\gamma(x)$ satisfies $f^{(\ell)}(x)>0$ for every $\ell$. Hence the forward finite differences $\Delta^\ell f(k)$ are positive for all $\ell$ and integer $k$
within the scope that they can be defined.}
For instance, their Lovelock invariants satisfy the following inequalities
\begin{align}
 \ell=0:\quad \gamma_k>1, (k\ge 1)\qquad  \left(\Leftrightarrow \quad 1>\frac{\LL_2}{R^2}>\frac{\LL_3}{R^3}>\frac{\LL_4}{R^4}>\cdots\ge 0 \right), 
\end{align}
\begin{align}
\ell=1:\quad  \gamma_{k+1}>\gamma_k,  (k\ge 0)\qquad  \left(\Leftrightarrow 
 1>\frac{\LL_2}{R^2}>\frac{\LL_3}{\LL_2 R}>\frac{\LL_4}{\LL_3 R}>\cdots\ge 0 \right),
\end{align}
and
\begin{align}
  \ell=2:\quad  \gamma_{k+1}\gamma_{k-1}>\gamma_k^2, (k\ge 1),
\end{align}
of which $k=1$ case gives $\gamma_2>\gamma_1^2$ since $\gamma_0=1$.
The question of 
\emph{whether the logarithm $\log \gamma_k$ is completely log-convex 
even in general CIRS cases} 
is an interesting problem, but in fact, as we will see in later sections, this question is also physically significant.
In this paper, verification is only possible for the first few terms of the sequences $\gamma_k$ with established formulas.

\begin{figure}[t]           
    \centering              
    \includegraphics[width=1.0\textwidth
    ]{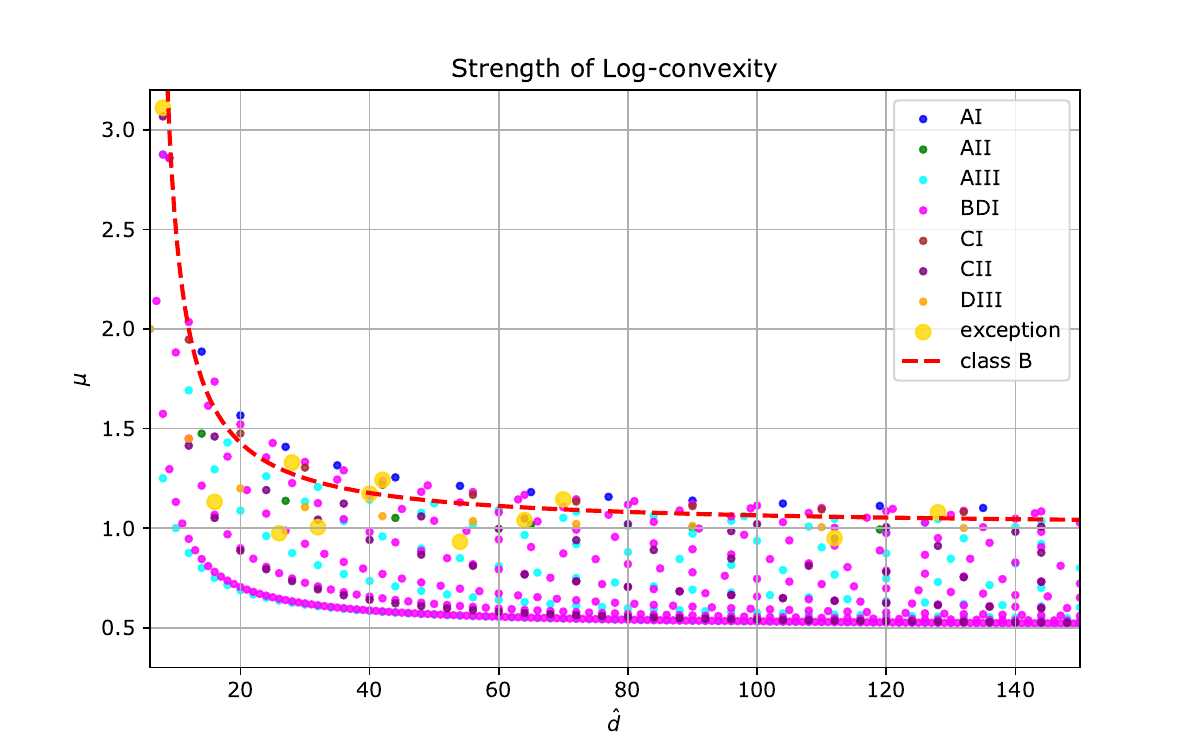} 
    \caption{$\hat d$ dependence of the log-convexity strength $\mu$.
    Data for $S^n$ and $\mathbb CP^m$ are located on the bottom layer of this figure.} 
    \label{fig:Convexity}      
\end{figure}

Using formulas \eqref{eq:lovelockR2}, \eqref{eq:LovelockR3} with appropriate assumptions on $(I_i, {\rm dim}\mathfrak h_i)$,
at least, one can show that
\begin{align}
    1>\frac{\LL_2}{R^2}>\frac{\LL_3}{R^3}\ge 0,
\end{align}
which implies $\gamma_1,\gamma_2>1$$(\ell=0)$.
However, to prove stronger inequalities, we need detailed information 
about $(I_i, {\rm dim}\mathfrak h_i)$ for each CIRS series.
For general CIRS spaces,
comprehensive direct verification using data listed in Table \ref{tab:Ihvaluelist} in addition to formulas \eqref{eq:lovelockR2}, \eqref{eq:LovelockR3} reveals that the following inequalities hold universally:
\begin{align}
    \gamma_2>\gamma_1>1,\qquad \gamma_2>\gamma_1^2,
\end{align}
where $\gamma_2>\gamma_1$ follows from the stronger inequality $\gamma_2>\gamma_1^2>\gamma_1$.
Specific proofs of $\gamma_2>\gamma_1^2$ are performed for each CIRS series; readers can verify them relatively easily and listing them all would be quite redundant. 
Therefore, we will not do so here and instead present all results in a graph.
Before this visualization, 
it is important to note that classical series allow discussion of dimensional dependence, and the dimensional dependence of $\gamma_1$ and $\gamma_2$ for large $\hat d$ in these series is universal,
\begin{align}
    \gamma_1=1+\frac{3}{\hat d}+{\mathcal O}(\hat d^{-3/2}),\qquad
      \gamma_2=1+\frac{6}{\hat d}+{\mathcal O}(\hat d^{-3/2}),
\end{align}
and largely independent of the type of CIRS series,
whereas ${\cal O}(\hat d^{-3/2})$ terms depend on them. 
 Furthermore, we observe the following universal behavior
\begin{align}
     \gamma_2-\gamma_1^2={\mathcal O}(\hat d ^{-2}),
\end{align}
which means unexpected cancellations of ${\cal O}(\hat d^{-3/2})$ terms.
From these observations, we find that 
it is convenient to define the following quantity
\begin{align}
\mu := \frac{\gamma_2-\gamma_1^2}{(\gamma_1-1)^2},\label{eq:mu}
\end{align}
which indicates a "strength" of the log-convexity
and will play important roles in later sections.
We plotted all data of $(\hat d,\mu)$ for CIRS spaces with $\hat d\le 150$ in Fig.\ref{fig:Convexity},
where one clearly observes that 
\begin{align}
    \mu > \frac12
\end{align}
 for all cases, at least up to $\hat d=150$.
To reinforce this result for finite $\hat d$, 
let us discuss the limit for large $\hat d$.
Taking the limit of large $\hat d$, we find that 
limiting values are universal  for AI, AII, DIII, CI series
and class B as
\begin{align}
    \lim_{n\to \infty}\mu=1,
\end{align}
and those for AIII, BDI and CII with keeping a ratio $r:=p/q$
also are universal,
\begin{align}
    \lim_{p,q\to \infty}\mu=\frac{1+3r+10r^2+3r^3+r^4}{2(1+2r+3r^2+2r^3+r^4)}=:\mu_\infty(r)\in \left[\frac12,1\right],
\end{align}
which takes the maximal value $\mu_\infty(1)=1$ again for $p=q$ case and 
the minimum value $\mu_\infty(0)=\mu_\infty(\infty)=1/2$ for cases 
where one of $p$ and $q$ is kept finite while taking the infinite limit of the other.
The mysterious universality in the above
suggests that 
at the limit of large $\hat d$, 
only the rank remains 
among the information contained in CIRS space.

Note that from Fig.\ref{fig:Convexity} and the above limits,
we observe that 
among CIRS spaces of the same dimension $\hat d$, 
rank-1 spaces, including $S^n$ and $\mathbb{C}P^m$, 
yield almost the smallest values of $\mu$.
These observations suggest some correlation between the rank of a CIRS space and the strength of log-convexity, and this fact will prove highly relevant to subsequent physical properties.
Furthermore, based on the fact that the series with the "weakest" log-convexity strength possesses complete log-convexity, we conjecture that \emph{all CIRS spaces may possess complete log-convexity}. This should be an interesting topic for future research.

\section{Metastable ghost-free Minkowski vacua}\label{sec:MetastableVacua}
\def\ul#1{{\underline {#1}}}
\subsection{Ansatz and effective 4D action}
Let us consider a Lovelock compactification on a general compact irreducible symmetric space (CIRS space).
In Section \ref{sec:irreduciblity}, the size modulus was treated as a constant parameter to establish the equivalence between the higher-dimensional equations of motion and the extremization of the moduli potential.
Here, by contrast, we promote the modulus to a dynamical scalar field and derive the four-dimensional effective action.

We consider the following ansatz:
a spacetime manifold is a direct product 
\begin{align}
    {\cal M}={\cal \ul M}\times {\cal \widehat M},\qquad 
    {\rm dim}({\cal \ul M})=4,\quad \hat d:={\rm dim}({\cal\widehat  M}),
     \quad (d=4+\hat d),
\end{align}
equipped with  a warped product metric\footnote{
Here we omit a Yang-Mills sector, just for simplicity.
By replacing 
\begin{align}
    dy^{\hat \mu}\quad \to \quad Dy^{\hat \mu}:=dy^{\hat \mu}-dx^{\ul \lambda} A_{\ul \lambda}^{\hat B}(x)\xi^{\hat \mu}_{\hat B}(y)
\end{align}
with using a Killing vector $\xi=T^{\hat B}\xi_{{\hat B}}^{\hat \mu}\p_{\hat \mu}$ and a gauge field $A=T_{\hat B}A^{\hat B}_{\ul \lambda}dx^{\ul \lambda}$ and $T_{\hat B}\in\mathfrak g$,  the metric becomes invariant under gauge transformation and 
thus, the effective action turns out to posses the Yang-Mills sector as a Kaluza-Klein zero mode. The ghost-free condition of that sector must also be checked, which is,
however, beyond the scope of this paper.
}
\begin{align}
    ds^2= \ul{g}_{\ul \mu\ul \nu}(x)dx^{\ul \mu}dx^{\ul \nu}
    +\hat \rho^2(x) \hat g_{\hat \mu \hat \nu}(y)dy^{\hat \mu}dy^{\hat \nu}.
\end{align}
Here $({\cal \ul M},\ul g)$ denotes the four-dimensional pseudo-Riemannian manifold describing our observable universe, while $({\cal\widehat  M},\hat g)$ is a CIRS space whose overall size is characterized by the size-modulus scalar field $\hat\rho(x)$.
In the following, we denote objects and indices related to the four-dimensional universe with an underline, and those related to the internal space with a hat.
Under this ansatz, the nonvanishing components of the Riemann curvature tensor are
\begin{align}
  R^{\ul \mu\ul \nu}{}_{\ul \rho \ul \sigma}(\ul g\oplus \hat \rho^2\hat g)&=R^{\ul \mu\ul \nu}{}_{\ul \rho \ul \sigma}(\ul g),\\
  R^{\hat \mu\hat \nu}{}_{\hat \rho \hat \sigma}(\ul g\oplus \hat \rho^2\hat g)
  &=\frac{1}{\hat \rho^{2} }\left(R^{\hat \mu\hat \nu}{}_{\hat \rho \hat \sigma}(\hat g)
  -2\delta^{\hat \mu}_{[\hat \rho}\delta^{\hat \nu}_{\hat \sigma]} {|\ul \nabla \hat \rho|^2}\right),\\
  R^{\ul \mu\hat \rho}{}_{\ul \nu \hat \sigma}(\ul g\oplus \hat \rho^2\hat g)&=
  -(\hat \rho^{-1}\ul \nabla^{\ul \mu}\ul \nabla_{\ul \nu}\hat \rho)\,\delta^{\hat \rho}_{\hat \sigma}.
\end{align}
Substituting this metric into the Lovelock action yields the effective 4D action for the graviton and the scalar $\hat\rho$ with higher-derivative correction terms. This effective theory is guaranteed to fall into the class of Horndeski theories \cite{Horndeski:1974wa,VanAcoleyen:2011mj}, since the Lovelock tensor remains second-order in derivatives and the scalar equation of motion is essentially determined by the trace of the higher-dimensional Lovelock tensor, ${\cal G}^{\hat \lambda}{}_{\hat \lambda}$.

Since we are interested in the low-energy effective theory, we assume that $\hat\rho$ varies slowly compared to the internal curvature scale.
Therefore, we truncate the derivative expansion at second order in $\nabla\hat\rho$.
 In this setting, each Lovelock invariant is expanded as,
\begin{align}
    &\LL_k(\ul g\oplus \hat \rho^2\hat g)\nn\\
    =&\hat \rho^{-2k}\LL_k(\hat g)
    +k \hat \rho^{-2(k-1)}\LL_{k-1}(\hat g)R(\ul g)+\frac{k(k-1)}{2}
    \hat \rho^{-2(k-2)}\LL_{k-2}(\hat g)\LL_2(\ul g)\nn\\
    &-k(\hat d-2k+2)\hat \rho^{-2(k-1)}\LL_{k-1}(\hat g)\left\{(\hat d-2k+1)\frac{|\ul \nabla \hat \rho|^2}{\hat \rho^2}+2\frac{\ul \Delta \hat \rho}{\hat \rho}\right\}\nn\\
    &-k(k-1)(\hat d-2k+4)\hat \rho^{-2(k-2)}\LL_{k-2}(\hat g)\left\{(\hat d-2k+3)R(\ul g)\frac{|\ul \nabla \hat \rho|^2}{\hat \rho^2}-4G^{\ul \mu \ul \nu}(\ul g)
    \frac{\ul \nabla_{\ul \mu} \ul \nabla_{\ul \nu}\hat \rho}{\hat \rho}\right\}\nn\\ 
    &-\frac{k(k-1)(k-3)}{2}(\hat d-2k+5)(\hat d-2k+6)
    \hat \rho^{-2(k-3)}\LL_{k-3}(\hat g)\LL_2(\ul g)\frac{|\ul \nabla \hat \rho|^2}{\hat \rho^2}\nn\\
    &+{\cal O}((\Delta \hat \rho, (\nabla \hat \rho)^2)^2).
\end{align}
Substituting the above result into the action we obtain 
\begin{align}
    S\Big|_{\rm ansatz}=\int_{\cal \ul M} d^4x\sqrt{-\ul g}
   & \left\{  -V_0(\hat \rho)+K(\hat \rho)R(\ul g)+J(\hat \rho)\LL_2(\ul g)\right.\nn\\
    &\quad-K''(\hat \rho)|\ul \nabla \hat \rho|^2-2 K'(\hat \rho)\ul \Delta \hat \rho\nn\\
    &\quad-2J''(\hat \rho)R(\ul g)|\ul \nabla \hat \rho|^2+8J'(\hat \rho) G^{\ul \mu\ul\nu}(\ul g)
    \ul \nabla_{\ul \mu} \ul \nabla_{\ul \nu}\hat \rho \nn \\
    &\quad \left.-3I''(\hat \rho)\LL_2(\ul g)|\ul \nabla \hat \rho|^2\quad +{\cal O}((\Delta \hat \rho, (\nabla \hat \rho)^2)^2)\quad \right\},
\end{align}
where $G^{\ul \mu\ul \nu}$ is the Einstein tensor and 
$V_0,K,J,I$ are polynomial functions with respect to $\hat \rho$, defined by,
\begin{align}
    V_0(\rho)&:=- \sum_{k=0}^{\lfloor d/2\rfloor}\hat \alpha_k \LL_k (\hat g) \rho^{\hat d-2k},  \\
    K(\rho)&:=\sum_{k=1}^{\lfloor d/2\rfloor}k\hat \alpha_k \LL_{k-1} (\hat g) \rho^{\hat d-2(k-1)}, \\
    J(\rho)&:=\sum_{k=2}^{\lfloor d/2\rfloor}\frac{k(k-1)}{2}\hat \alpha_k \LL_{k-2} (\hat g) \rho^{\hat d-2(k-2)},\\
    I(\rho)&:=\sum_{k=3}^{\lfloor d/2\rfloor}\frac{k(k-1)(k-2)}{6}\hat \alpha_k \LL_{k-3} (\hat g) \rho^{\hat d-2(k-3)},
\end{align}
with parameters $\hat \alpha_k:=\alpha_k\,{\rm Vol}(\hat {\cal M},\hat g )$ rescaled by the internal-volume factor,
\begin{align}
    {\rm Vol}(\hat {\cal M},\hat g ):=\int_{\cal \hat M}d^{\hat d}y\sqrt{\hat g}.
\end{align}

Here, the effective Gauss-Bonnet term $\LL_2(\ul g)$ for the four-dimensional graviton 
and the scalar field $\hat \rho$ are non-trivially coupled through the function $J(\hat \rho)$. As originally anticipated, the singular behavior of black holes and the early universe inevitably excites the scalar field due to the presence of this term. While terms involving $J, I$ will be important in future research, for now,
in this paper, these terms are not essential for establishing the healthiness of the vacuum as a preliminary step.
\subsection{Conditions for (meta)stable ghost-free Minkowski vacuum}

As a limit of the time evolution, we expect that the manifold of our universe becomes a maximally symmetric space as
\begin{align}
 \lim_{t\to \infty}   R^{\ul \mu\ul \nu}{}_{\ul \rho \ul \sigma}(\ul g)=2H^2\delta^{\ul \mu}_{[\ul \rho}\delta^{\ul \nu}_{\ul \sigma]} 
\end{align}
with the Hubble parameter $H$.  There, with constant $\hat \rho=\hat \rho_0,H$,
the equations of motions ${\cal G}^{\mu \nu}=0$ reduce to 
\begin{align}
       V_0(\hat \rho_0)=6 H^2 K(\hat \rho_0),\quad 
    V_0'(\hat \rho_0)=12H^2 K'(\hat \rho_0)+24H^4 J'(\hat \rho_0)
\end{align}
which are just extremization conditions of an effective "potential"
given by  
\begin{align}
    V_{\rm eff}(\hat \rho,H)=\frac{V_0(\hat \rho)}{H^4}-12\frac{K(\hat \rho)}{H^2}-24J(\hat \rho).
\end{align}
At such a vacuum point satisfying these equations,
any viable model must be ghost-free and stable.

Our primary objective is to establish stable, ghost-free models and their parameter settings in the Minkowski vacuum limit $H\to 0$.
Since viable models are expected to admit a smooth Minkowski limit
to be taken continuously, we consider the method of obtaining finite $H$ cases by parameter perturbations from this limit to be the most natural approach.

Let us consider the Minkowski vacuum satisfying the extremization condition
\begin{align}
    V_0(\hat \rho_0)=0,\quad V_0'(\hat \rho_0)=0,\label{eq:Minkowski}
\end{align}
where $V_0(\hat \rho)$ is nothing but a potential for the scalar and thus 
stability of the scalar requires $V_0''(\hat \rho_0)>0$. 
The coefficient of the Einstein-Hilbert term $R(\ul g)$ for the 4D graviton must be positive
and determines the effective Einstein constant as
\begin{align}
   \frac{1}{2\kappa_4}:=K(\hat \rho_0).
\end{align}
The positivity of $K(\hat\rho_0)$ ensures that the 4D graviton has the correct sign kinetic term.

It is also necessary to verify that the scalar field is not a ghost, but to correctly handle the kinetic term of the scalar field,
one must consider the contribution of the scalar component of the gravitational field.
For this purpose, transitioning from the Jordan frame to the Einstein frame is most convenient.
Let us take a conformal transformation $\ul g\to e^{2\varphi}\ul g$ 
so that the coefficient is fixed to be constant as
\begin{align}
    e^{2\varphi}K(\hat \rho)={\rm const.}=\frac1{2\kappa_4}\quad \Leftrightarrow \quad \varphi=-\frac12\ln\left(2\kappa_4K(\hat \rho)\right),
\end{align}
where the normalization is chosen such that 
$\varphi=0$ at the vacuum point $\hat \rho =\hat \rho_0$.

Using the following formula
\footnote{The formula for a Riemann curvature tensor is given as
\begin{align}
   R^{\ul \mu\ul \nu}{}_{\ul \rho \ul \sigma}(e^{2\phi}\ul g)=
   e^{-2\phi}\left(R^{\ul \mu\ul \nu}{}_{\ul \rho \ul \sigma}(\ul g)+4\delta^{[\ul \mu}_{[\ul \rho}Q^{\ul \nu]}_{\ul \sigma]} \right),\quad Q^{\ul \mu}_{\ul \nu}:=-\ul \nabla^{\ul \mu}\ul \nabla_{\ul \nu}\phi+
   \ul \nabla^{\ul \mu}\phi \ul \nabla_{\ul \nu}\phi-\frac12 \delta^{\ul \mu}_{\ul \nu}|\ul\nabla\phi|^2.
\end{align}
}
\begin{align}
       e^{2\varphi} R(e^{2\varphi}\ul g)=R(\ul g)-6\left(\ul \Delta \varphi 
       +|\ul \nabla \varphi|^2\right),
\end{align}
with $\varphi$ given above,
finally we obtain the low-energy effective 4D action around the Minkowski vacuum as
\begin{align}
    S_{\rm eff}=\int_{\cal \ul M}d^4x \sqrt{-\ul g}\left\{\frac{1}{2\kappa_4}R(\ul g)
    -\frac12 {\cal N}(\hat \rho)\ul \nabla^{\ul \mu}\hat \rho \ul \nabla_{\ul \mu}\hat \rho -V(\hat \rho)\right\}
\end{align}
with 
\begin{align}
    {\cal N}( \rho):=\frac{1}{\kappa_4}\left(\frac32 \left(\frac{K'(\rho)}{K(\rho)}\right)^2-\frac{K''(\rho)}{K(\rho)}\right),\qquad V(\hat \rho):=\frac{V_0(\hat \rho)}{(2\kappa_4K(\hat \rho))^2}. \label{eq:KineticCoeff}
\end{align}
The condition ${\cal N}(\hat \rho_0) > 0$ guarantees that the scalar field is not a ghost, taking into account its mixing with the gravitational degrees of freedom.
Note that while the potential is rescaled, the location and existence of the Minkowski vacuum remain unchanged as long as $K(\hat{\rho}_0) > 0$.
Here we omit the higher-curvature corrections proportional to $I(\hat \rho),J(\hat \rho)$ 
for describing low-energy dynamics around the Minkowski vacuum.

In summary,
at the Minkowski-vacuum point satisfying \eqref{eq:Minkowski}, a healthy theory requires, at least, the following local conditions
\begin{align}
  \hat \alpha_1>0,\quad V_0''(\hat \rho_0)>0,  \quad K(\hat \rho_0)>0,\quad {\cal N}(\hat \rho_0)>0.
\end{align}
In the following, we discuss these local conditions with presenting explicit models.
The global structure of the $\hat\rho$ moduli space and its global healthiness will be discussed in Sec.~\ref{sec:GlobalStability}.
\subsection{No-go theorem for Einstein--Gauss--Bonnet gravity}

We consider the theory consisting of the Einstein--Hilbert term together with a single higher-order Lovelock invariant,
\begin{align}
    S = \int_{\cal M} d^d x \,\sqrt{-g} \left(\alpha_0 + \alpha_1 R + \alpha_m \LL_m \right), 
    \qquad 2 \le m \le \frac{\hat d}{2}.
\end{align}
Here $m=2$ corresponds to Einstein--Gauss--Bonnet (EGB) gravity with a cosmological term, 
while $m>2$ incorporates higher-order Lovelock corrections.

Compactifying on a CIRS space $\hat{\cal M}$ with scale modulus $\rho$, 
the effective potential and kinetic coefficient are
\begin{align}
    V_0(\rho) &= - \hat \alpha_0 \rho^{\hat d} - \hat \alpha_1 R(\hat g)\rho^{\hat d-2} - \hat \alpha_m \LL_m(\hat g) \rho^{\hat d-2m},\\
    K(\rho) &= \hat \alpha_1 \rho^{\hat d} + m \hat \alpha_m \LL_{m-1}(\hat g) \rho^{\hat d-2(m-1)}.
\end{align}

Tuning one parameter so that the potential admits a four-dimensional Minkowski vacuum at $\rho = \hat \rho_0$, we fix
\begin{align}
    \hat \alpha_0 &= -\frac{m-1}{m} \,\hat \alpha_1 R(\hat g)\, \hat \rho_0^{-2}, \qquad
    \hat \alpha_m = -\frac{\hat \alpha_1}{m} \frac{R(\hat g)}{\LL_m(\hat g)} \hat \rho_0^{2(m-1)}.
\end{align}
This ensures
\begin{align}
    V_0''(\hat \rho_0) = 4 (m-1) \hat \alpha_1 R(\hat g) \hat \rho_0^{\hat d-4} > 0,
\end{align}
so the vacuum is locally stable at the level of the potential.

The kinetic coefficient at the vacuum becomes
\begin{align}
    K(\hat \rho_0) = \hat \alpha_1 \bigl(1 - \gamma_{m-1}\bigr) \hat \rho_0^{\hat d},
\end{align}
where $\gamma_k=\gamma_k(\hat{\cal M})$ is the ratio of Lovelock invariants introduced in the previous section.  
However, as discussed in Sec.~\ref{sec:log-convexity}, we conjecture that 
\begin{align}
    \gamma_{m-1}(\hat{\cal M}) > 1
\end{align}
holds for any Lovelock order $m$ and any CIRS space $\hat{\cal M}$, 
with explicit verification in the following cases:
\begin{enumerate}
    \item arbitrary $\hat{\cal M}$ with $m=2$ (Gauss--Bonnet term) or $m=3$ (Cubic Lovelock term),
    \item $\hat{\cal M} = S^n$ or $\mathbb CP^n$ for arbitrary $m \in [2, \lfloor \hat d/2 \rfloor]$.
\end{enumerate}
For these cases we obtain $K(\hat \rho_0) < 0$, establishing a no-go theorem: 
Einstein--Gauss--Bonnet gravity, and more generally any model with only a single Lovelock term beyond Einstein--Hilbert, 
cannot provide a ghost-free Minkowski vacuum on these internal spaces.

This result, particularly for \(m=2\), extends earlier observations of ghost
and instability problems in Gauss--Bonnet compactifications—previously
identified mainly for spherical internal spaces—to the case of arbitrary
CIRS spaces (see, e.g.,
\cite{Calcagni:2006ye,Charmousis:2008ce,Pavluchenko:2015daa,DeFelice2024}.)
Our formulation reveals the underlying geometric mechanism behind the no-go theorem: universal log-convexity concerning the Lovelock invariants.

An important dimensional consequence follows: since the $m$-th Lovelock invariant $\LL_m$ vanishes identically for dimensions $\hat{d} < 2m$, higher-order terms ($m \ge 3$) can only contribute non-trivially if the internal dimension satisfies $\hat{d} \ge 6$. For lower-dimensional CIRS spaces, such as
\begin{align}
S^n \quad (n \le 5), \qquad
\mathbb CP^{2}, \qquad
\frac{SU(3)}{SO(3)},
\end{align}
the action is kinematically restricted to the Einstein--Gauss--Bonnet form. In these cases, the no-go theorem becomes an absolute barrier that cannot be circumvented 
by simply adding a higher-order term from the Lovelock series.
These spaces are thus fundamentally unsuitable as internal manifolds for a ghost-free Minkowski vacuum.

Some studies in the literature simplify analysis by reducing the number of terms, but our results for $m\ge 3$ extend the no-go theorem to higher-order Lovelock terms, compelling improvements in much of those works.
The next subsection shows that escaping the no-go theorem requires increasing complexity by adding more terms, not by raising their order.


\subsection{Four-term model:  ghost-free and locally-stable vacua}
Since the three-term (Einstein–Gauss–Bonnet) model inevitably contains ghosts, we now introduce a four-term model by adding the cubic Lovelock term, whose action is given by
\begin{align}
    S=\int_{\cal M}d^dx \sqrt{-g}\left(\alpha_0+\alpha_1 R+\alpha_2 \LL_2+\alpha_3 \LL_3\right), \quad \hat d\ge 6\label{eq:4term}
\end{align}
which provides an additional degree of freedom that can potentially cure this instability.
Indeed, previous studies have indicated that higher-order terms can expand the parameter space for stable compactification \cite{Chirkov:2018xrd}.
In this subsection, we demonstrate that the cubic term is indeed essential for reconciling the ghost-free condition with the linear stability of the Minkowski vacuum.

In this model $V_0$ and $K$ are given by the following polynomials:
\begin{align}
    V_0(\rho)&=-\hat \alpha_0\rho^{\hat d}-\hat \alpha_1R(\hat g)\rho^{\hat d-2}-\hat \alpha_2
    \LL_2(\hat g)\rho^{\hat d-4}-\hat \alpha_3
    \LL_3(\hat g)\rho^{\hat d-6},\\
    K(\rho)&=\hat \alpha_1\rho^{\hat d}+2\hat \alpha_2
    R(\hat g)\rho^{\hat d-2}+3\hat \alpha_3
    \LL_{2}(\hat g)\rho^{\hat d-4}.
\end{align}
Here we assume $\hat d\ge 6$ so that $\LL_{1,2,3}(\hat g)>0$.
Tuning one parameter within $\{\hat \alpha_i\}$, we require the following form of the potential:
\begin{align}
     V_0(\rho)=\lambda \rho^{\hat d-6}(\rho^2-\hat \rho^2_0)^2(\rho^2-c\,\hat \rho^2_0),
\end{align}
with introducing new parameters $\lambda,\hat \rho_0, c$, so that it allows for a 4D-flat-spacetime vacuum at $\rho=\hat \rho_0$. In terms of these new parameters, $\{\hat \alpha_i\}$ are rewritten as
\begin{align}
    \hat \alpha_0=-\lambda,\quad \hat \alpha_1=\frac{(2+c)}{R(\hat g)}\lambda \hat \rho^2_0,\quad
    \hat \alpha_2=-\frac{(1+2c)}{\LL_2(\hat g)}\lambda \hat \rho^4_0,\quad
    \hat \alpha_3=\frac{c}{\LL_3(\hat g)}\lambda \hat \rho^6_0.
\end{align}
Setting $c=0$, this model reduces to the three-term model.
Here we require inequalities, $\hat \alpha_1>0$ and $V''_0(\hat \rho_0)>0$ which give
\begin{align}
    (2+c)\lambda>0, \quad (1-c)\lambda>0, \quad \Rightarrow\quad 1>c>-2,\quad\lambda>0.
\end{align}

Next, we examine the ghost-free condition at the vacuum point $\hat\rho=\hat\rho_{0}$,
under the parameter region satisfying the above inequalities.
Substituting the above parameterization into $K(\hat \rho_0)$, we obtain
\begin{align}
    K(\hat \rho_0)=\frac{\lambda}{R(\hat g)}\left\{(1-4\gamma_1+3\gamma_2)c-2(\gamma_1-1)\right\}\hat \rho^{\hat d+2}_0.
\end{align}  
Here $\gamma_1$ and $\gamma_2$ are the dimensionless ratios of the Lovelock invariants introduced in Sec.~\ref{sec:log-convexity}.
The universal log-convexity, $\gamma_2>\gamma_1^2>\gamma_1>1$, discussed there, give the following inequalities
\begin{align}
    K(\hat \rho_0)\Big|_{c=1}\propto & \,3(1+\gamma_2-2\gamma_1)=3(1+\mu)(\gamma_1-1)^2>0,\\
    K(\hat \rho_0)\Big|_{c=0}\propto &-2(\gamma_1-1)<0,
\end{align}
where $\mu>0$ is the "strength" of the log-convexity defined in Eq.\eqref{eq:mu}.
We thus find that the model admits a tensor-sector ghost-free window for the parameter $c$:
\begin{align}
    K(\hat \rho_0)>0 \quad {\rm for~} 1>c>c_{\rm inf},\quad
    c_{\rm inf}:=\frac{2}{2+3(1+\mu)(\gamma_1-1)},
\end{align}
where $K(\hat \rho_0)$ vanishes at $c=c_{\rm inf}$.
Furthermore, if  $c$ satisfies $c>c_{\rm inf}$ while taking a value sufficiently close to $c_{\rm inf}$,
then the $(K'/K)^2$ term is dominant\footnote{
Strictly speaking, this statement is not always true.
This is true if and only if 
\begin{align}
 0\not = K'(\hat\rho_0)\big|_{c=c_{\rm inf}}\propto
    (\gamma_1-1)c_{\rm inf}\times( (\mu-1)-(\gamma_1-1)(\mu+1)).
\end{align}
Surprisingly, as will be seen in the next section, all class-B cases provide counterexamples of this assumption as
$K'(\hat\rho_0)=0$ at $c=c_{\rm inf}$.
However, since these cases conversely demonstrate 
${\cal N}(\hat\rho_0)> 0$ at $c=1$, the final conclusion remains essentially unchanged.
} in the definition of ${\cal N}(\hat \rho)$ given in \eqref{eq:KineticCoeff}, and thus ${\cal N}(\hat \rho_0)$ is positive definite:
\begin{align}
    \exists \tilde c\in (c_{\rm inf},1],\quad \forall c\in (c_{\rm inf},\tilde c): \quad {\cal N}(\hat \rho_0)>0,
\end{align}
which guarantees ghost-free in the scalar sector.
Therefore, \emph{the four-term model with any CIRS space
provides a locally-stable and ghost-free Minkowski vacuum
if the parameter is correctly chosen.}

\subsection{Robust AdS preference}
Moving beyond the local analysis of the previous subsection, we now examine the global structure of the moduli space in the four-term model. A crucial observation is that a region with negative potential energy always coexists with the Minkowski vacuum:
\begin{align}
  \forall c\in (c_{\rm inf},1)\subset (0,1),\quad \forall \rho \in (0,\sqrt{c}\hat \rho_0):\qquad  V_0(\rho)<0.
\end{align}
According to the Coleman-De Luccia (CDL) formalism \cite{Coleman:1980aw}, the existence of such a deeper AdS vacuum\footnote{
Strictly speaking, the effective potential $V(\hat{\rho})$ is constructed under the assumption that a Minkowski vacuum exists at $\hat{\rho} = \hat{\rho}_0$. Thus, a local minimum in the $V < 0$ region does not necessarily correspond to a precise AdS vacuum solution of the full theory. Nevertheless, the existence of a region where $V < 0$ serves as a clear indicator of the Minkowski vacuum's metastability, as it signals the presence of lower-energy configurations in the moduli space.}
is a dangerous sign. In such a landscape, the Minkowski vacuum is at best metastable, and one must strictly evaluate its decay rate to ensure the longevity of the universe.

This issue appears particularly severe in the ghost-free parameter regime, where the unintended AdS region often manifests "dangerously close" to the Minkowski vacuum in terms of the potential landscape. One might hope that further extending the theory could eliminate this instability. Recalling that the inclusion of the cubic Lovelock term allowed us to bypass the no-go theorem of the three-term (EGB) model, a natural expectation would be that adding a quartic Lovelock term (the five-term model) might finally eliminate the unintended AdS vacuum.

While the quartic Lovelock invariants for general CIRS are not yet available, we can explicitly test this possibility for specific cases such as $S^n$ and $\mathbb{C}P^m$. However, as detailed in Appendix~\ref{sec:RobustAdS}, the result is negative: the AdS vacuum persists even in the five-term model. This behavior stems from a higher-level "log-convexity" of the Lovelock invariants. We conjecture that this property holds for general CIRS, which possess even stronger log-convexity, suggesting that the AdS vacuum cannot be avoided simply by adding more higher-order terms. We refer to this pathological tendency as the \emph{"Robust AdS preference."}

One might then consider if quantum corrections could lift the potential to remove these vacua. However, such large quantum effects would necessitate a complete reformulation of the vacuum stability analysis, which lies beyond the scope of this paper. In this work, we assume the validity of the effective theory and seek a classical resolution.

This leads us to a crucial re-evaluation of the "proximity" between vacua. A naive assessment based solely on the potential's shape ignores the underlying geometry of the moduli space—specifically, the kinetic structure of the scalar field. To evaluate the true physical stability, one must look beyond the coordinate distance on the potential and instead analyze the geodesic distance on the moduli space. If this distance is finite, the CDL tunneling probability remains non-zero, posing a persistent threat.

This observation naturally sets the stage for the next section. We will demonstrate that while the AdS preference is indeed robust against adding higher-order terms, the \emph{"Kinetic Barrier" mechanism}—available uniquely in higher-rank spaces—can effectively isolate the Minkowski vacuum by pushing the unintended AdS region to the boundary of the moduli space
located at infinite geodesic distance.

\section{Effectively stable vacua via kinetic barrier}\label{sec:GlobalStability}
In the previous section, we examined the local positivity of the kinetic coefficient $K(\hat\rho)$ that controls the graviton and moduli dynamics, focusing only on its value at $\hat\rho = \hat\rho_0$, which realizes the flat 4D spacetime.
 Here, we shall investigate the global behavior of $K(\hat \rho)$ over the entire domain of the moduli space. In general, within the physically admissible range of $\hat \rho$, there can appear regions where $K(\hat \rho)$ becomes negative, leading to the emergence of gravitational ghosts.  

Should such behavior be regarded as pathological and thus excluded from the theory?  
At first sight, one might think so, since $K<0$ leads to a ghost-like kinetic term.  
Paradoxically, this apparent pathology turns out to play a crucial role in promoting the  metastable vacuum to be stable: the vanishing of $K(\hat\rho)$ introduces a dynamical barrier that isolates different regions of moduli space, as we shall see below.

\subsection{Kinetic barrier mechanism}
Recall that the coefficient ${\cal N}$ of the kinetic term for the size-modulus scalar field $\hat \rho$ is given in terms of the polynomial function $K(\hat \rho)$ as in Eq.\,\eqref{eq:KineticCoeff}. If there exists a region ($\hat \rho<\hat \rho_*$) where $K(\hat \rho)<0$, then $K(\hat \rho)$ must vanish linearly at its boundary $\hat \rho=\hat \rho_*$ as 
\begin{align}
    K(\hat \rho) \sim a(\hat \rho-\hat \rho_*),
\end{align}
where $a>0$ because $K(\hat\rho)$ must increase toward the healthy region.
According to Eq.\,\eqref{eq:KineticCoeff}, the kinetic term then exhibits the following singular behavior:
\begin{align}
    \frac{1}{2}{\cal N}(\hat \rho)(\partial_\mu \hat \rho)^2
    \sim \frac{3a^2}{4\kappa_4}\frac{(\partial_\mu \hat \rho)^2}{(\hat \rho-\hat \rho_*)^2}.
\end{align}
To remove this coordinate singularity, we perform a field redefinition $\hat \rho \mapsto \phi$ so that the kinetic term in the healthy domain $\hat \rho>\hat \rho_*$ becomes canonical and 
$\phi$ reflects the geodesic distance:
\begin{align}
    \hat \rho = \hat \rho_* + b\, e^{a'\phi},
    \qquad
    a' = \sqrt{\frac{2\kappa_4}3}\frac{1}{a}, 
    \quad b>0.
\end{align}
In this canonical coordinate system, the boundary of the ghost region $(\hat \rho<\hat \rho_*)$ is mapped to $\phi=-\infty$ and can never be reached.  
Therefore, the ghost modes associated with that region are physically irrelevant.

This mechanism can be applied to the problem of avoiding the robust AdS preference. Even if a deeper "true vacuum" exists, its transition can be blocked by a ghost region positioned between the Minkowski vacuum and the deeper vacuum. In such a configuration, the ghost region acts as a kinetic barrier, making the Minkowski vacuum effectively behave as a true vacuum.

\begin{figure}[h]
\centering
\includegraphics[width=0.6\textwidth,trim=50 0 50 0, clip]{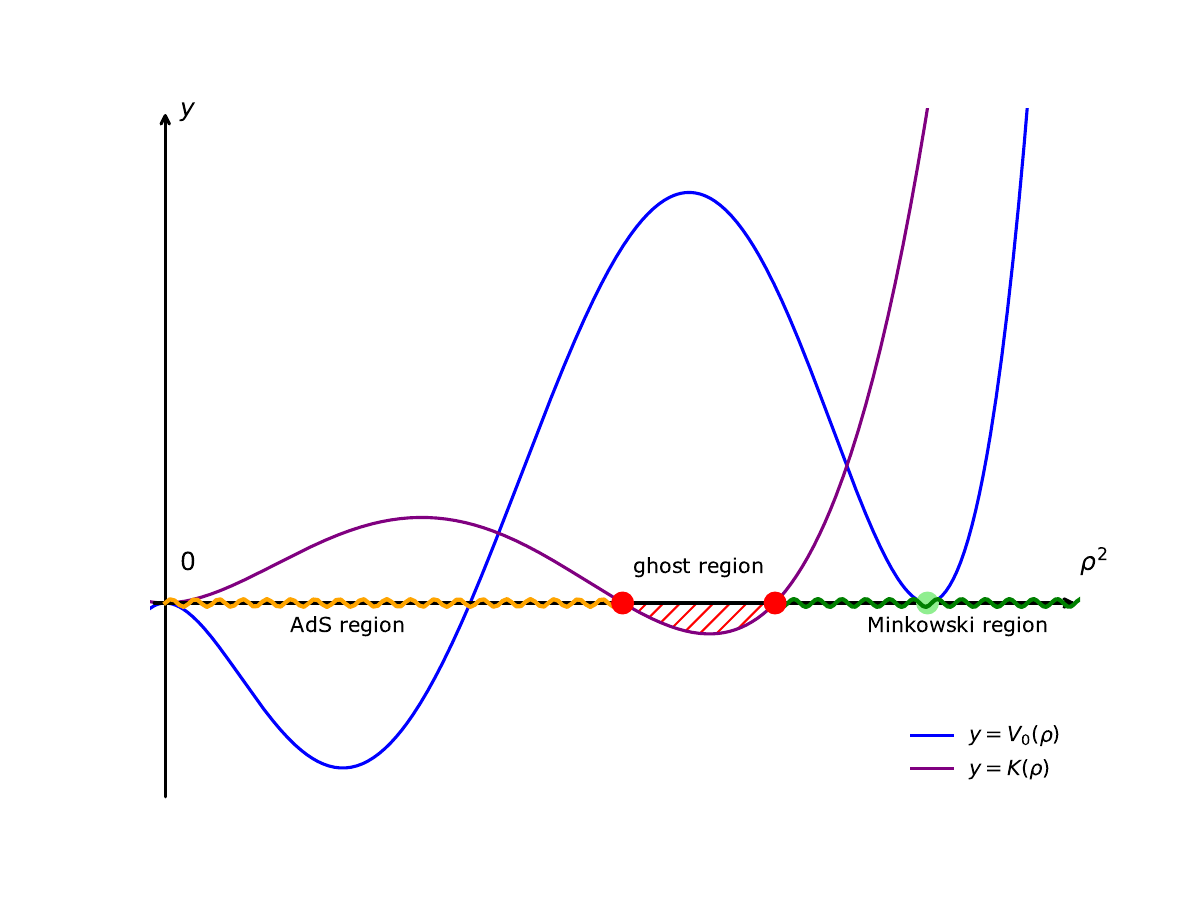}
\caption{Schematic illustration of the kinetic barrier mechanism.
The moduli space is divided into three regions: the Minkowski region(green), the ghost region(red), and the AdS region(orange). The wavy lines indicate the physically accessible domains. The divergence of the kinetic term at the boundaries of the ghost region creates a barrier that geodetically isolates the Minkowski vacuum from the AdS region.}
\label{fig:KineticBarrier}
\end{figure}

The global structure of this mechanism is schematically illustrated in Fig.~\ref{fig:KineticBarrier}. We categorize the moduli space into three distinct regions: the healthy Minkowski region, where our vacuum resides; the ghost region ($K < 0$), which acts as the kinetic barrier; and the AdS region, which contains a deeper vacuum.

As shown in the figure, the divergence of the kinetic term at the boundary ($\hat{\rho} = \hat{\rho}_*$) ensures that the Minkowski and AdS regions are geodetically disconnected. This separation physically traps the system within the Minkowski region, thereby circumventing the "Robust AdS Preference" by rendering the unintended AdS vacuum unreachable. Whether such a configuration is actually realized in Lovelock compactification depends on the specific form of the invariants, which we shall examine in the following subsections.

\subsection{Stable window in the four-term model}
Now, let us use the four-term model given in Eq.\,\eqref{eq:4term}, discussed in the previous section, to examine whether this mechanism indeed operates.  
What we must confirm is the existence of a boundary $\hat \rho=\hat\rho_*$ of the ghost region that satisfies the following conditions:
\begin{align}
   \exists \hat \rho_*:\quad K(\hat \rho_*)=0,
   \qquad 
   \sqrt{c}\,\hat \rho_0 < \hat \rho_* < \hat\rho_0.
\end{align}

We define a function $f(x;c)$ by
\begin{align}
    K(\hat \rho_0 \sqrt{x}) &\propto \quad f(x;c)
    := (2+c)x^2 - 2(2c+1)\gamma_1 x + 3c\gamma_2,
\end{align}
which is a quadratic function of $x(>0)$ and takes its minimum at
\begin{align}
   x_{\mathrm{m}}(c) := \frac{2c+1}{2+c}\gamma_1.
\end{align}
Since $\gamma_2>\gamma_1^2>\gamma_1>1$ holds for any CIRS space, we find
\begin{align}
   f(c;c)>0,
   \qquad 
   x_{\mathrm{m}}(c)>c
   \quad \text{for } c\in(0,1),
\end{align}
and hence, if $f(x;c)=0$ has real roots, they must satisfy $x>c$.

Moreover, the ghost-free condition requires a lower bound $c_{\mathrm{inf}}$ such that
\begin{align}
    f(1;c)>0 
    \quad \text{for } c>c_{\mathrm{inf}},
    \qquad 
    \exists\, c_{\mathrm{inf}}: f(1;c_{\mathrm{inf}})=0.
\end{align}
Note that $f(x;c)$ is linear in $c$, and that $f(x;1)\ge 3(\gamma_2-\gamma_1^2)>0$.  
Therefore, there always exists $c_{\rm sup}=c_{\rm sup}(\gamma_1,\gamma_2)\in(c_{\mathrm{inf}},1)$ such that
$f(x_{\mathrm{m}}(c_{\rm sup});c_{\rm sup})=0$.  
For $c\in(c_{\mathrm{inf}},c_{\rm sup})$, there appears a ghost region $(x^-_*,x^+_*)$ satisfying
\begin{align}
    f(x;c)<0,
    \qquad  
    x\in(x^-_*,x^+_*),
\end{align}
where
\begin{itemize}
    \item $(x^-_*,x^+_*)\subset (c,1)$ if $x_{\mathrm{m}}(c_{\mathrm{inf}})<1$ $\left(\Leftrightarrow \frac{\p f}{\p x}(1;c_{\mathrm{inf}})>0\right)$,
    \item $(x^-_*,x^+_*)\subset (1,\infty)$ if $x_{\mathrm{m}}(c_{\mathrm{inf}})>1$  $\left(\Leftrightarrow \frac{\p f}{\p x}(1;c_{\mathrm{inf}})<0\right)$,
\end{itemize}
and the critical case $x_{\mathrm{m}}(c_{\mathrm{inf}})=1$ corresponds to
$c_{\mathrm{inf}}=c_{\rm sup}$, $x^-_*=x^+_*=1$, and $\frac{\p f}{\p x}(1;c_{\rm inf})=0$.  
The cases satisfying $x_{\mathrm{m}}(c_{\mathrm{inf}})<1$ are precisely those we seek, 
since in these cases the ghost region $(x^-_*,x^+_*)$ lies between the flat vacuum ($x=1$) and 
the region ($x<c$) where the potential is negative.

Combining the above properties, we find that the condition for the existence of such a stable window can be written compactly as follows:
there exists a stable window $(c_{\rm inf}, c_{\rm sup})$ for the parameter $c$,
which is strictly contained within the ghost-free window $(c_{\rm inf}, 1)$, given by
\begin{align}
    c_{\rm inf}=\frac{2}{2+3(1+\mu)(\gamma_1-1)},\quad c_{\rm sup}=\frac{1}{1+3\mu'+3\sqrt{\mu'(1+\mu')}} \label{eq:csupcinf}
\end{align}
with 
\begin{align}
    \mu := \frac{\gamma_2-\gamma_1^2}{(\gamma_1-1)^2} >0,\quad 
    \mu':= \frac{\gamma_2-\gamma_1^2}{\gamma_1^2}>0,
\end{align}
if and only if a CIRS space satisfies
\begin{align}
    \gamma_1-1<\frac{\mu-1}{\mu+1}.  \label{eq:stability}
\end{align}
We call such a CIRS space \emph{stabilizable}.
For the four-term model with such a stabilizable space as the internal space,
by choosing the parameter $c$ within this stable window,
the desired Minkowski vacuum becomes effectively stable due to the kinetic barrier.

The kinetic barrier mechanism provides an additional advantage.
In the region where $x>x_*^{\pm}>0$, the coefficient for the kinetic term for the scalar field $\hat \rho$ is automatically positive definite:
\begin{align}
    {\cal N}(\hat \rho_0 \sqrt{x})\propto&\frac{(\hat d-2)(\hat d-4)}{2x}
    +\frac{2(\hat d-5)}{x-x_*^+}+\frac{2(\hat d-5)}{x-x_*^-}\nonumber\\
    &+\frac{6x}{(x-x_*^+)^2}+\frac{6x}{(x-x_*^-)^2}
    +\frac{4x}{(x-x_*^+)(x-x_*^-)}>0,\quad(\hat d\ge 6).
\end{align}
Therefore, once Inequality \eqref{eq:stability} is satisfied, the effective theory for the graviton and the scalar field is completely healthy within this stable window at the classical level.


\subsection{Stabilizable CIRS spaces: classification and examples}
\begin{figure}[h]
\centering
\includegraphics[width=1.0\textwidth,trim=50 0 50 0, clip]{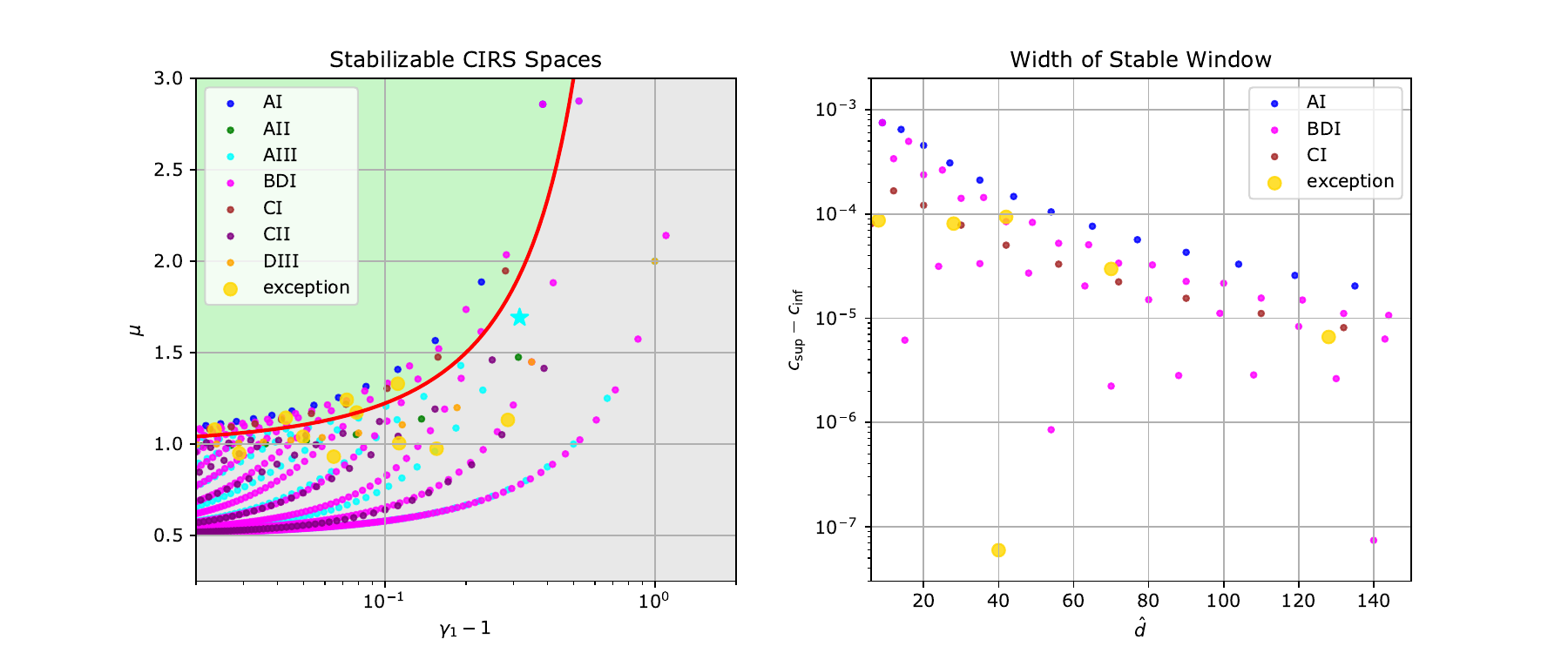}
\caption{Stabilizable spaces in the allowed region (light green) are shown in the left panel, and the width of the stable window for each case is displayed in the right panel.}
\label{fig:StableWindow}
\end{figure}

Here, we classify CIRS spaces based on whether they can be stabilized in the four-term model.
In the left panel of Fig.\ref{fig:StableWindow}, we show all data points $(\gamma_1,\mu)$ for CIRS spaces with $\hat d \le 150$, together with the allowed region satisfying Inequality\eqref{eq:stability}, shown in light green.
The condition is rather stringent and rules out most symmetric spaces. Nevertheless, an infinite number of candidates survive.

The stabilizable CIRS spaces are:
\begin{itemize}
\item \textbf{Classical series:} \quad AI ($n\ge 4$), BDI ($p\simeq q\ge 2$), CI ($n\ge 2$)
\item \textbf{Exceptional series:} \quad EI, EII, EV, EVIII, FI, G.
\end{itemize}
These stabilizable CIRS spaces are heavily concentrated in a narrow band very close to the boundary of the allowed region. Since the stability window collapses to zero width precisely on the boundary, this observed concentration implies that all stability windows $(c_{\rm inf}, c_{\rm sup})$ for these stabilizable CIRS spaces are extremely narrow, as shown in the right panel of Fig.~\ref{fig:StableWindow}.
The origin of this rather striking concentration pattern—which resembles the way settlements tend to cluster along a riverbank—is not yet well understood.

Even more remarkably, all group-type (class B) cases—namely the compact simple Lie groups equipped with the standard bi-invariant metric—appear exactly on the stability boundary
\begin{align}
\gamma_1 - 1
= \frac{3}{\hat d - 3}
= \frac{\mu - 1}{\mu + 1},
\end{align}
although the origin of this relation also remains unclear.

Crucially, {\it all rank-1 spaces, including $S^n$ and $\mathbb{C}P^m$, are excluded}.
In the figure, they lie on the opposite side of the stability boundary from the allowed region, and within that domain they align along its outer edge, forming a dense cluster far from the boundary itself.
This reflects the fact that their log-convexity strength, measured by $\mu$, is far too weak to approach the stability condition.
While the correlation between rank and log-convexity strength is not exact across different dimensions, it is observed that $\mu$ generally increases with the rank for a fixed dimension. This trend explains why only higher-rank spaces possess sufficient log-convexity to approach and cross the stability boundary.

These observations are supported by the large-$\hat d$ behavior:
\begin{align}
    \lim_{\hat d\to \infty}\frac{1}{\gamma_1-1}\frac{\mu-1}{\mu+1}=\left\{ 
    \begin{array}{cc}
       \infty  &  \hbox{for AI, BDI($p=q$),CI},\\
        1 & \hbox{for class B},\\
        \frac{25}{27}& \hbox{for AIII($p=q$)},\\
        -\infty & \hbox{for AII, DIII,CII($p=q$)},
    \end{array}\right.
\end{align}
where "$\infty$" denotes divergence of order $\sqrt{\hat d}$. 
For AIII, BDI, and CII with $p\not =q$ we observe that this limit gives negative divergence since $\lim_{\hat d\to \infty}\mu<1$.
Therefore, for stabilizable CIRS spaces, $\mu$ behaves
\begin{align}
    \mu-1={\cal O}(\hat d^{-1/2}),\quad \text{with } \gamma_1=1+\frac{3}{\hat d}+{\cal O}(\hat d^{-3/2}) \label{eq:mubehavior}
\end{align}
which implies that the width of the stability window given in Eq.\eqref{eq:csupcinf} behaves 
\begin{align}
    c_{\rm sup}-c_{\rm inf}={\cal O}(\hat d^{-2}),
    \quad \text{with }c_{\rm inf}=1-\frac{9}{\hat d}+{\cal O}(\hat d^{-3/2}).\label{eq:width}
\end{align}
Here, the terms up to order $\hat d^{-3/2}$ in $c_{\rm inf}$ and $c_{\rm sup}$ given in \eqref{eq:csupcinf} cancel each other out exactly in this difference.
 
The first few examples for such stable ones are 
\begin{align}
    &\frac{SO(5)}{SO(3)\times SO(2)}\simeq \frac{Sp(2)}{U(2)}\,(\hat d=6),\quad \frac{G_2}{SO(4)}\,(\hat d=8),\quad \frac{SO(6)}{SO(3)\times SO(3)}\simeq \frac{SU(4)}{SO(4)}\,(\hat d=9),\nonumber\\
    &\frac{SO(7)}{SO(4)\times SO(3)}\,(\hat d=12),\quad \frac{Sp(3)}{U(3)}\,(\hat d=12),
    \quad \frac{SU(5)}{SO(5)}(\hat d=14),\dots\,.\label{eq:sCIRS}
\end{align}
Among the AIII series, the space
\begin{align}
\frac{SU(5)}{U(1)\times SU(2)\times SU(3)},(\hat d=12)
\end{align}
is a familiar example in contexts such as symmetry breaking patterns of grand unified models.
In the present setup it narrowly fails to satisfy the stability condition,
\begin{align}
\frac{1}{\gamma_1-1}\frac{\mu-1}{\mu+1}
=\frac{83}{102}<1, \quad {\rm with~}\gamma_1=\frac{25}{19},\quad
\gamma_2=\frac{19}{10},
\end{align}
and is therefore excluded at cubic order. For reference, this AIII point is marked by a star in Fig.~\ref{fig:StableWindow}.
This exclusion simply reflects the criterion adopted here; higher-order Lovelock terms could in principle alter the conclusion.

We note that the situation differs between six-dimensional CIRS spaces and those of higher dimensions.
There are three six-dimensional CIRS spaces: in addition to the first one listed in \eqref{eq:sCIRS}, there are $S^6$ and $\mathbb CP^3$, which cannot be stabilized in the four-term model. Unlike spaces with $\hat d\ge 8$, even when higher-order Lovelock terms are added in cases of the $\hat d=6$ spaces, these higher-order terms do not affect the vacuum structure. Consequently, the result that $S^6$ and $\mathbb CP^3$ cannot be stabilized remains unchanged.
\subsection{Mass hierarchy and phenomenological constraints}
\begin{figure}[h]
\centering
\includegraphics[width=\textwidth]{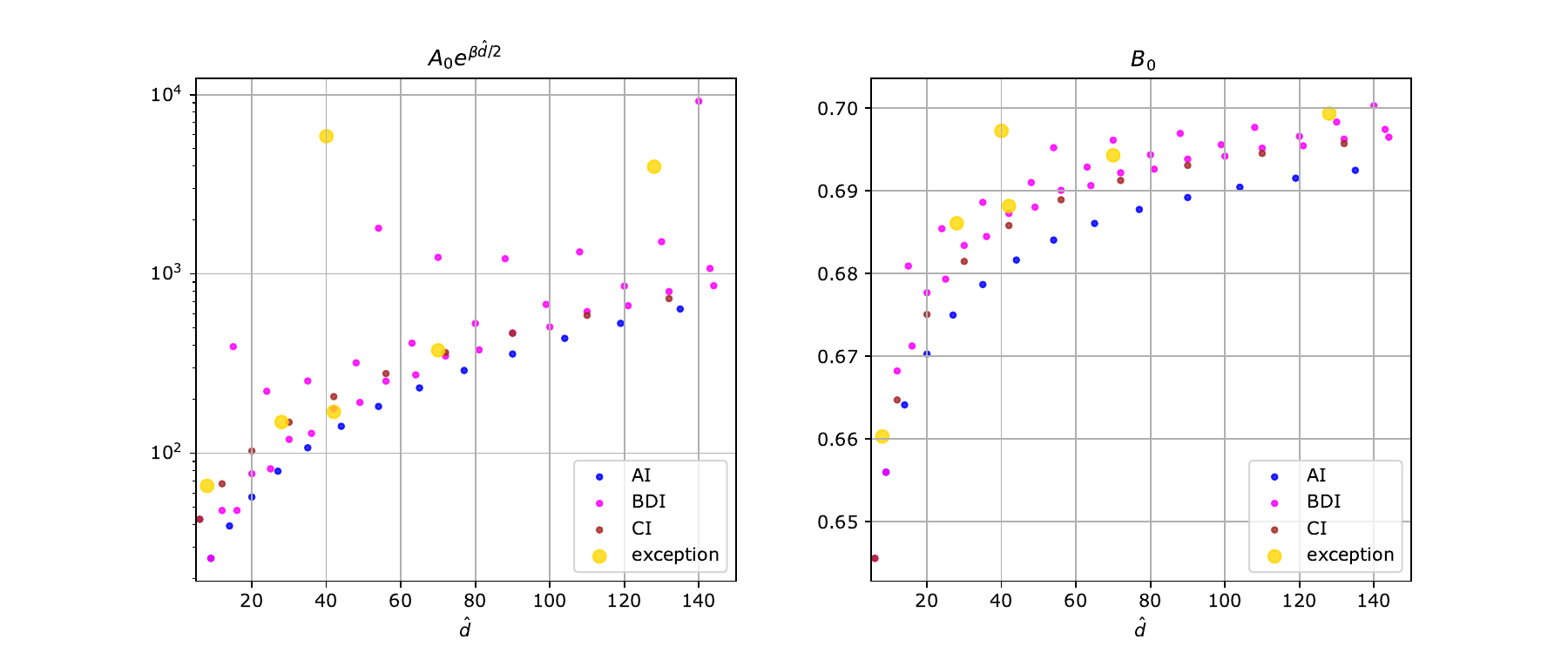}
\caption{
Coefficients $A_0({\cal \hat M})e^{\beta \hat d/2}$ (left) and 
$B_0({\cal \hat M})$ (right) for stabilizable CIRS spaces, plotted as functions of 
the internal dimension $\hat d$ for various classical series.}
\label{fig:MassRatio}
\end{figure}

So far, we have established that the four-term model can satisfy the local healthiness requirements for admitting Minkowski spacetime as a vacuum.
Here we examine whether there exist parameter choices compatible with current collider constraints within this framework.
After fixing one parameter to obtain the Minkowski vacuum, the four-term model contains three parameters $\lambda, \hat\rho_0$, and $c$. The overall coefficient $\lambda$ and the internal radius $\hat\rho_0$ can be rewritten as\footnote{Here we use the fact that all stabilizable CIRS spaces belong to class A, where the Ricci scalar satisfies
 $R(\hat g)=N_{\mathfrak g}({\rm ad})\,\hat d$.}
\begin{align}
\lambda=\frac{\Lambda_d}{\kappa_d},\qquad 
    \hat \rho_0=\sqrt{\frac{R(\hat g) }{2(2+c)\Lambda_d}}
    =\sqrt{\frac{N_{\mathfrak g}({\rm ad})\hat d }{2(2+c)\Lambda_d}}
\end{align}
where $\Lambda_d$ is the cosmological constant and $\kappa_d$ the Einstein constant.
Using these parameters, we introduce the following four phenomenologically relevant mass scales: the reduced Planck masses in the original Lovelock dimension and in four dimensions, respectively,
\begin{align}
    M_{{\rm Ll}}:=\kappa_d^{-\frac1{d-2}}=(2\alpha_1)^{\frac{1}{\hat d+2}},\qquad
    M_{{\rm Pl}}:=\kappa_4^{-\frac12}=\sqrt{2K(\hat \rho_0)},
\end{align}
together with the mass of the size modulus $\hat\rho$ and the expected KK mass scale determined by the internal radius,
\begin{align}
    M_{\hat \rho}:=\sqrt{ \frac{V_0''(\hat \rho_0)}{{\cal N}(\hat \rho_0)}},\qquad
    M_{\rm KK}:=\sqrt{\frac{2(2+c)\Lambda_d}{\hat d}}.
\end{align}
Here we removed the factor $N_{\mathfrak g}({\rm ad})$ from the definition of $M_{\rm KK}$ to make it independent of normalization.
For a given value of $c$, these four masses satisfy two independent relations:
 \begin{align}
        \frac{M_{\rm Ll}}{M_{{\rm Pl}}}
    =A(\hat {\cal M},c)\left(\frac{M_{\rm KK}}{M_{{\rm Ll}}}\right)^{\frac{\hat d}2}, \qquad  \frac{M_{\hat \rho}}{M_{\rm KK}} =B(\hat {\cal M},c),\label{eq:massratios}
 \end{align}
 for certain coefficients $A(\hat {\cal M},c),B(\hat {\cal M},c)$.
 Recall that $K(\hat \rho_0)$ vanishes at $c=c_{\rm inf}$, 
 and that the stable window $(c_{\rm inf},c_{\rm sup})$ is extremely narrow.
 Consequently, $K(\hat \rho_0)$ and ${\cal N}^{-1}(\hat \rho_0)$ behave linearly with respect to the parameter 
 \begin{align}
    s:=\frac{c-c_{\rm inf}}{c_{\rm sup}-c_{\rm inf}}\in(0,1),\qquad c=s \,c_{\rm sup}+(1-s) \, c_{\rm inf},
 \end{align}
 whereas the remaining quantities are essentially constant in $s$.
These features imply the following simple dependence:
 \begin{align}
    A(\hat {\cal M},c)\approx \frac{ A_0(\hat {\cal M})}{\sqrt{s}},\qquad
B(\hat {\cal M},c)\approx B_0(\hat {\cal M})\sqrt{s},
 \end{align}
 where the coefficients $A_0$ and $B_0$, specific to each CIRS space, are
 \begin{align}
     A_0(\hat {\cal M}) 
     =\sqrt{\frac{(2+c_{\rm inf})c_{\rm inf}}{2(c_{\rm sup}-c_{\rm inf})}\times 
    \frac1{  {\rm vol}({\cal \hat M}) }},
 \end{align}
 \begin{align}
     B_0({\cal \hat M})=\sqrt{\frac{2(1+\mu)(c_{\rm sup}-c_{\rm inf})\hat d}{9c_{\rm inf}^2\left((\mu-1)-(\mu+1)(\gamma_1-1)\right)^2}}.
 \end{align}
Here ${\rm vol}({\cal \hat M})$  is the normalization-independent volume,
\begin{align}
    {\rm vol}({\cal \hat M}):=&\int_{\cal \hat M}d^{\hat d}y\sqrt{\hat g}\left(\frac{R(\hat g)}{\hat d}\right)^{\frac{\hat d}{2}}\\
    =&{\rm Vol}({\cal \hat M},\hat g)[N_{\mathfrak g}({\rm ad})]^{\frac{\hat d}{2} }
    ={\rm Vol}({\cal \hat M},N_{\mathfrak g}({\rm ad})^{-1}\hat g).
\end{align}
The volumes of all CIRS spaces were exhaustively computed by Abe and Yokota \cite{AbeYokota1997}, who used the Killing form as normalization, corresponding to $N_{\mathfrak g}({\rm ad})=1/2$ for class-A spaces.
From the explicit formulas for the CIRS spaces that admit stabilization, one observes an almost universal behavior in their $\hat d$-dependence. In particular, all stabilizable classical series approach a common limiting value\footnote{For all rank-1 spaces, including $S^n$ and $\mathbb{C}P^m$,
this limiting value is $(\log(2\pi)+1)/2$ which is 7\% larger than that of the stabilizable spaces.}
\begin{align}
   \lim_{\hat d\to \infty}\frac1{\hat d}\log {\rm vol}({\cal \hat M})=\frac12\log \pi +\frac34=:\beta.
\end{align}
For later reference, we note that $e^{\beta/2}\simeq 10^{0.287}$.
Therefore a factor $e^{-\beta \hat d/2}$ gives the dominant contribution in $A_0({\cal \hat M})$, deriving a factor contributing to a large hierarchy on the masses for large $\hat d$.
Using the Abe--Yokota formulas \footnote{
For the CI series, the original Abe--Yokota formula contains a factor 
$\pi^{n(n+2)/2}$, which leads to half-integer powers of $\pi$ in odd dimensions.
This behavior is inconsistent with the volume formulas of other classical series
and, in particular, fails to match the BDI series in the six-dimensional case 
($p=3,q=2$), where the two spaces are isometric.
We therefore replace this factor by $\pi^{n(n+1)/2}$, which restores consistency
among classical series and yields the expected dimensional dependence of the volume.},
we plotted a subleading factor $A_0({\cal \hat M})e^{\beta \hat d/2}$ in the left panel of Fig.~\ref{fig:MassRatio}.
On the other hand, according to \eqref{eq:mubehavior} and \eqref{eq:width}, one finds that the coefficients $B_0({\cal \hat M})$ for the classical series approach to certain constants in the large-$\hat d$ limit, 
and using further careful calculations 
the coefficients again turn out to approach a common limiting value,
\begin{align}
     \lim_{\hat d\to \infty}B_0({\cal \hat M})=
    \lim_{\hat d\to \infty} \sqrt{\frac{4(c_{\rm sup}-c_{\rm inf})\hat d}{9\left(\mu-1\right)^2}}
     =\frac1{\sqrt{2}}.
\end{align}
We plotted $B_0({\cal \hat M})$ in the right panel of Fig.~\ref{fig:MassRatio}.
In both of the above two limits, we observe universal values independent of the classical series, whereas the reason is unclear.
From these universal properties, it can be seen that the coefficients 
$A_0({\cal \hat M}),B_0({\cal \hat M})$ are almost entirely determined by the dimensions of a CIRS space alone, whereas one observe fine structures from Fig.~\ref{fig:MassRatio}.

Using \eqref{eq:massratios}, assuming ratios between the three masses $M_{\rm Ll},M_{\rm KK}$ and $M_{\hat \rho}$ determines their relative values with respect to the reduced Planck mass $M_{\rm Pl}=2.44\times 10^{18}\mathrm{GeV}$.
Note that the factor $(M_{\rm KK}/M_{\rm Ll})^{\hat d/2}$ in \eqref{eq:massratios} causes a large hierarchy between the planck mass and the others.
Here, let us assume, for example, the following physically reasonable ratios 
\begin{align}
    M_{\rm KK}=0.1 M_{\rm Ll},\qquad M_{\hat\rho}\approx 0.2 M_{\rm KK}=0.02 M_{\rm Ll}
\end{align}
with setting $s=0.1$.
\begin{figure}[h]
\centering
\includegraphics[width=0.9\textwidth]{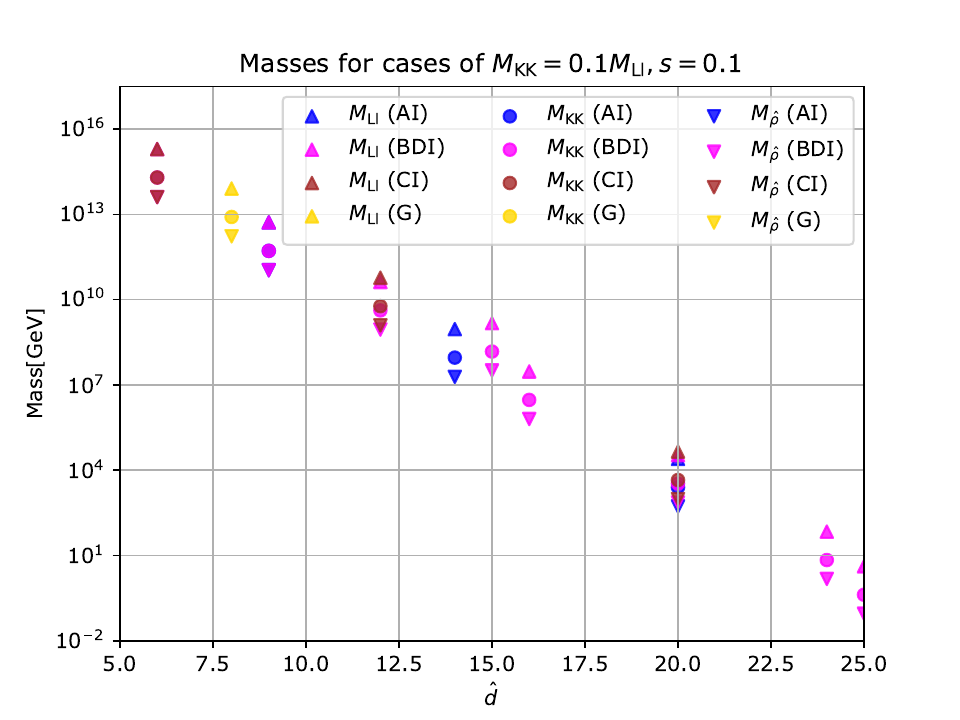}
\caption{Mass scales as functions of the internal dimension $\hat d$ for stabilizable CIRS spaces, assuming $M_{\rm KK}=0.1 M_{\rm Ll}$ and $s=0.1$.}
\label{fig:Masses}
\end{figure}
Figure~\ref{fig:Masses} plots the three masses of the stabilizable CIRS spaces under these assumptions, focusing on those roughly at the TeV scale or above.
There, it becomes clear that as the dimension increases, the masses decrease exponentially.
Theoretically, the set of stabilizable CIRS spaces contains infinite sequences, but considering consistency with experiments, high-dimensional CIRS space is rejected. Under the above assumptions, $\hat d$ must be approximately 20 or less.
Even under the most lenient assumption of $M_{\rm KK}=M_{\rm Pl}$ and $s=1$,
$\hat d$ should be restricted to be less than around 65, and dimensions larger than this should be phenomenologically excluded in the four-term model.

From the viewpoint of superstring theory, internal dimensions larger than 
$\hat d = 6$ are difficult to justify. Among the stabilizable CIRS spaces, 
the case $\hat d = 6$, corresponding to
\begin{align}
    {\cal \hat M} = \frac{SO(5)}{SO(3)\times SO(2)},
\end{align}
is particularly well motivated.
Although the hierarchy generated in this case is much weaker, one may still 
bring the Kaluza--Klein scale down to the TeV range by choosing 
$M_{\rm KK}/M_{\rm Ll} = \mathcal{O}(10^{-4})$.
Such a parameter choice would place the lightest KK modes within the reach of 
current collider experiments.
We do not pursue this possibility further in this work.

\subsection{Comments on de Sitter spacetime}

So far, we have focused on the Minkowski spacetime limit obtained by tuning one of the parameters of the theory.
This choice is motivated by the fact that local properties, such as vacuum stability and the absence of ghosts,
are governed by energy scales far above the observed cosmological constant and are therefore insensitive
to whether the background spacetime is exactly Minkowski or slightly de Sitter.
Nevertheless, applying the present model to cosmology is clearly important.
Before concluding this section, we therefore briefly outline how a small nonzero cosmological constant may be incorporated.

To this end, we slightly shift the parameter $\alpha_0$ from its Minkowski value to $(1+\epsilon)\alpha_0$,
which lifts the potential as
\begin{align}
    V_0(\hat \rho)\ \longrightarrow\ 
    V_\epsilon(\hat \rho)=V_0(\hat \rho)+\epsilon \lambda \hat \rho^{\hat d},
\end{align}
where $\epsilon$ is an extremely small parameter.
Introducing the Hubble parameter $H$ and assuming an exponentially stable de Sitter spacetime with constant $H$,
the equations of motion reduce to
\begin{align}
       V_\epsilon(\hat \rho_\epsilon)=6 H^2 K(\hat \rho_\epsilon),\quad 
    V_\epsilon'(\hat \rho_\epsilon)=12H^2 K'(\hat \rho_\epsilon)+24H^4 J'(\hat \rho_\epsilon),
\end{align}
where the vacuum expectation value of $\hat \rho$ is slightly shifted to $\hat \rho_\epsilon$.
Since the scale of $H$ is extremely small compared to all other mass scales in the problem,
the following approximations hold:
\begin{align}
    \epsilon\approx \frac{12H^2K(\hat \rho_0)}{\lambda \hat \rho_0^{\hat d}}
    \approx 24\frac{s(c_{\rm sup}-c_{\rm inf})}{c_{\rm inf}\hat d}\left(\frac{H}{M_{\rm KK}}\right)^2,
\end{align}
and
\begin{align}
    \hat \rho_\epsilon -\hat \rho_0\approx \frac{6H^2}{V_0''(\hat \rho_0)}
    \left(2 K'(\hat \rho_0)-\frac{\hat d}{\hat \rho_0}K(\hat \rho_0)\right)\approx {\cal O}\left[\left(\frac{H}{M_{\rm KK}}\right)^2\right]\times \hat \rho_0.
\end{align}
This shift is negligibly small, and there is therefore no need to reconsider the vacuum stability or the ghost-free condition. Indeed, since the Minkowski vacuum has been proven to be locally stable and ghost-free with a finite mass gap, a sufficiently small deformation to de Sitter spacetime $(H \ll M_{\text{KK}})$ cannot trigger any linear instability. Moreover, the presence of the expansion naturally provides Hubble friction, further ensuring the exponential stability of the vacuum against small fluctuations. Thus, we can safely conclude that the health of our compactified vacuum remains perfectly intact in a realistic cosmological setting.

We note that the ratio $H/M_{\rm KK}$ appears instead of $H/M_{\rm Pl}$ in the above expressions,
which may be interpreted as a mild alleviation of the hierarchy problem,
corresponding to an effective UV cutoff at the KK scale rather than the Planck scale.
In practice, however, this effect is negligible, and $\epsilon$ still parametrizes
the enormous hierarchy between cosmological and microscopic scales.
A detailed discussion of the cosmological constant problem is beyond the scope of this paper.

\paragraph{Remark}
It is worth emphasizing that the four-term model studied in this paper
should be interpreted with some care.
For the case $\hat d = 6$, the existence of a healthy Minkowski vacuum
requires only the tuning of the effective cosmological constant.
However, for $\hat d > 6$, the analysis is further restricted to
a subspace of the full parameter space,
not by physical necessity but by technical limitations of the model.

Unless such restrictions are justified by an underlying symmetry
or a fundamental principle,
it would be inappropriate to regard special choices of parameters
together with a purely classical analysis
as a resolution of the hierarchy problem.
The purpose of this work is therefore modest:
we have merely demonstrated that there exist regions in parameter space
where Minkowski spacetime arises as a consistent and healthy vacuum solution,
and we do not claim to address deeper issues such as the origin of
hierarchies in nature.

Similar claims have appeared in the literature based on specific
choices of Lovelock couplings and compactification schemes,
but our analysis suggests that such conclusions should be interpreted
with caution.


\section{Summary and Discussion}\label{sec:summary}
\paragraph{Summary}

In this paper, we investigated the healthiness of effective theories obtained by compactifying higher-dimensional Lovelock gravity on CIRS spaces, requiring that the four-dimensional Minkowski spacetime be a vacuum of the theory. We first showed that any three-term model, including the Einstein–Gauss–Bonnet case, inevitably suffers from a ghostlike graviton. In contrast, in the four-term model with the cubic Lovelock term included, there always exists a parameter region free of such ghosts; however, we found that Minkowski vacua in this region are generically metastable across the class of CIRS spaces considered, being accompanied by energetically favored Anti–de Sitter vacua. We refer to this generic and universal feature as the Robust AdS preference.

To overcome this obstruction, we proposed the kinetic barrier mechanism and demonstrated that it operates only for a subset—though an infinite family—of CIRS spaces. We further identified the extremely narrow parameter region in which this mechanism becomes effective. Finally, among the infinitely many ghost-free models, we argued that only finitely many can actually yield realistic mass parameters.

The novelty of this work lies in identifying the Robust AdS preference as a structural obstruction inherent to Lovelock compactifications, and in proposing the kinetic barrier mechanism as a concrete and dynamical way to evade it. Crucially, we showed that this mechanism does not operate on the spherical internal spaces commonly studied in the literature and instead requires CIRS spaces of sufficiently high rank.

\paragraph{Limitations of the Present Analysis and Required Future Checks}

It should be emphasized that the healthiness examined in this paper concerns only the low-energy effective theory of four-dimensional gravity coupled to the modulus of the internal radius. A complete analysis must ensure the absence of both ghosts and tachyons for all fluctuations around the Minkowski vacuum. In particular, it is known that a Yang–Mills sector inevitably appears as KK zero modes; if those modes turn out to be ghostlike, then even the four-term model would be invalidated, and one would be forced to introduce the fifth or higher Lovelock terms. Owing to technical difficulties, we have left these analyses for future work, but they represent essential steps toward establishing the full healthiness of the theory. Moreover, the study of the mass spectrum and the emergent Yang–Mills sector is not only mandatory but also of significant interest from the viewpoint of potential applications.

\paragraph{Beyond the Four-Term Model and the Kinetic Barrier}

The kinetic barrier mechanism proposed in this work excludes not only the spherical series but also phenomenologically interesting families such as the AIII series and the class-B spaces (simple Lie groups). However, this conclusion applies only to the four-term model; once fifth or higher Lovelock terms are included, the outcome may change. It should also be emphasized that the kinetic barrier is not necessarily the unique way to prevent Minkowski spacetime from being destabilized by competing vacua. Even if certain internal spaces are excluded by our criterion, being sufficiently close to the critical boundary would imply that the quantum tunneling rate is extremely suppressed, so that the lifetime of our Minkowski vacuum could easily exceed the age of the universe. Moreover, quantum effects not considered in this paper may lift the potential near the origin and thereby restore stability. These directions also offer interesting possibilities for future investigation.

\paragraph{Implications for Cosmology and Black-Hole Physics}

It should be emphasized that the present analysis concerns the low-energy effective theory around four-dimensional Minkowski vacua.
In regimes involving strong gravity, such as the early universe or black-hole spacetimes, one must return to the full higher-dimensional Lovelock equations.
In this sense, our results do not diminish the significance of existing dynamical and numerical studies.
Rather, the vacua and parameter regions identified here provide natural and physically well-motivated starting points for such analyses.

\paragraph{Interpretation: Robust AdS Preference and the Role of the Kinetic Barrier}
In Lovelock compactifications, the realization of a Minkowski vacuum typically requires parameter tuning and is accompanied by regions of negative vacuum energy in the moduli space. This structural tendency implies that a healthy Minkowski vacuum is generically metastable due to possible tunneling toward an AdS vacuum.

Within the class of models analyzed in this work, we have shown that this pathology can be cured when the moduli space is dynamically disconnected by a kinetic barrier. In such cases, the singular locus is sent to infinite moduli-space distance, causing the action of any tunneling trajectory toward the AdS region to diverge. As a result, semiclassical decay of the Minkowski vacuum is prevented.

Although our explicit analysis is restricted to the four-term model, the origin of the Robust AdS preference appears to rely on generic features of moduli potentials rather than on details of Lovelock gravity. It is therefore natural to expect that similar pathologies may persist in higher-order Lovelock theories, and possibly in a broader class of gravitational effective theories. From this viewpoint, the kinetic barrier mechanism can be regarded as a prototypical “cure” for a recurrent structural problem.

\paragraph{Speculative Interpretation: Kinetic Barrier and a Generalized Cosmic Censorship}

A personal motivation for studying Lovelock theory has been the expectation that higher–curvature terms—especially those involving higher powers of the Riemann tensor—make the theory intrinsically sensitive to the formation of spacetime singularities through its equations of motion. In this broader sense, 
one may heuristically view the theory as suggesting a generalized form
of the cosmic censorship idea: the dynamics themselves might prevent the system from evolving toward pathological geometries.

Although the present work concerns a problem that precedes cosmology or black-hole physics—namely, the stabilization of Minkowski space—it appears that the model already exhibits a primitive form of such a censoring mechanism. The kinetic-barrier phenomenon described here shows that once a healthy vacuum is realized, dynamical trajectories cannot approach regions of field space where the gravitational sector becomes ghost-like. In this way, the theory seems to enforce its own healthiness by dynamically excluding pathological domains.

The analysis in this paper is entirely classical. For any realistic application, quantum corrections are unavoidable, and it is natural to worry that the existence conditions of the healthy vacuum might be destabilized by such corrections. Whether an analogue of the kinetic barrier persists in a full quantum treatment remains an important direction for future work.

\paragraph{Higher-order Lovelock terms and the robustness of the AdS preference}

In this work, the Robust AdS preference has been established rigorously within the four-term model, and explicitly verified for the five-term model in simple internal spaces such as $S^n$ and $\mathbb{C}P^m$.
Whether this preference persists when arbitrarily higher-order Lovelock terms are included remains an open question.

For general CIRS spaces, addressing this problem is currently obstructed by the lack of explicit expressions for higher-order Lovelock invariants. However, for highly symmetric spaces such as $S^n$ and $\mathbb{C}P^m$, these invariants are known in closed form, and the problem reduces to an algebraic analysis of the two polynomials $V_0(\rho)$ and $K(\rho)$.

We therefore expect that a systematic study of higher-order Lovelock models on these spaces could decisively clarify whether the Robust AdS preference is a universal feature of Lovelock compactifications, or merely an artifact of truncation at low order. We leave this important question for future work.



\paragraph{Universal Properties and Open Mysteries}

The "universal log-convexity" employed in this study was confirmed only for the first few terms of the Lovelock invariants. Whether this property holds universally for higher-order $k$ and higher level $\ell$ across all classes of CIRS spaces remains an open question. Proving the generality of this mathematical property would be a pursuit of pure mathematical interest, and exploring its full extent in the context of physical stability would be equally significant.

In fact, throughout this study, we repeatedly encountered unexpected universalities. The most mysterious of these is the fact that CIRS in Class B align precisely on the critical curve of the stability window. This mathematical feature manifests physically as the extreme proximity of stable CIRS data to the critical boundary, which consequently results in an exceptionally narrow allowed region in the parameter space. Whether these phenomena are rooted in a deep, underlying principle that renders them inevitable, or are merely a series of mathematical coincidences, remains an absolute mystery without a single clue.
\appendix

\section{ Evaluation of the invariants $I_i(\mathfrak{m})$}\label{sec:calcI}
In this appendix, we clarify the group-theoretic conventions used in Table~\ref{tab:Ihvaluelist} and describe the algebraic procedures employed to determine the index ratios \(I_i\).
\subsection{Conventions and Definitions}
\paragraph{Definition of $I_i$:} For a given CIRS space $G/H$, the isotropy representation $\mathfrak{m}$ is decomposed into irreducible representations of the sub-algebras $\mathfrak{h}_i$ of the stabilizer $\mathfrak{h} = \bigoplus_i \mathfrak{h}_i$. The index ratio $I_i$ is defined by the ratio of the normalization constants (Dynkin indices) of the representation $\mathfrak{m}$ and the adjoint representation of $\mathfrak{h}_i$:
\begin{align}
I_i := I_{\mathfrak{h}_i}(\mathfrak{m}) = \frac{N_{\mathfrak{h}_i}(\mathfrak{m})}{N_{\mathfrak{h}_i}(\text{ad})}.
\end{align}
If $\mathfrak{h}$ contains an Abelian factor $\mathfrak{z}$, the corresponding $N_{\mathfrak{h}_i}(\text{ad})$ vanishes, and we formally treat $I_i$ as $\infty$. In practical calculations, this can be understood as a limit where the structure constants of the non-Abelian part dominate.

\paragraph{Notation of Representations:} Throughout the table, bold numbers (e.g., ${\bf n}$, ${\bf 26}$) denote the representation with that specific dimension. For classical groups, ${\bf n}$ typically refers to the (minimal) fundamental  representation.

\begin{itemize}
    \item $\text{Sym}^k$: The k-th symmetric power of the representation.
    \item $\Lambda^k$: The k-th alternating (anti-symmetric) power.
\item Subscript 0 (e.g., $\text{Sym}^2_0$): This indicates the trace-free part of the representation, where the trace with respect to the invariant tensor (e.g., the metric) has been removed.
\item {\bf spin}: The irreducible spinor representation of $\mathfrak{so}(n)$.
For odd $n=2k+1$, its dimension is $2^{k}$.  
For even $n=2k$, it denotes a half-spinor (chiral) representation with dimension $2^{k-1}$.
\end{itemize}
\paragraph{Real and Complex Representations:}  The direct sum $\mathfrak{m} \simeq R \oplus \text{c.c.}$ is used when the representation $\mathfrak{m}$ is a real representation formed by a complex representation $R$ and its complex conjugate. In this case, the index $I_i$ for $\mathfrak{m}$ is simply twice that of the single complex representation $R$.
\begin{itemize}
    \item The subscript "+" (e.g., $R_+$) denotes that the representation carries a specific $U(1)$ charge under the $\mathfrak{z}$ factor of $\mathfrak{h}$.
    \item If a representation is complex but not explicitly written as a sum with its complex conjugate, it is understood that a reality condition (such as self-duality) has been imposed. 
\end{itemize}
\paragraph{Exceptional Groups:} The bold numbers ${\bf 26, 27}$, and ${\bf 56}$ refer to the minimal fundamental representations of $\mathfrak{f}_4$, $\mathfrak{e}_6$, and $\mathfrak{e}_7$, respectively. 

\subsection{Algebraic Reduction and Derivation Methods}
The index ratios $I_i$ are defined as the ratio of two traces within the same algebra $\mathfrak{h}_i$. As such, they are normalization-independent; any overall scaling of the generators cancels out. This ensures that the values in Table~\ref{tab:Ihvaluelist} are purely algebraic invariants of the representation $\mathfrak{m}$, allowing them to be evaluated using standard results in representation theory without detailed knowledge of the Cartan decomposition $\mathfrak{g} = \mathfrak{h} \oplus \mathfrak{m}$.

For a CIRS space where the stabilizer is a direct sum $\mathfrak{h} = \mathfrak{h}_1 \oplus \mathfrak{h}_2$, the isotropy representation often decomposes as a tensor product $\mathfrak{m} \simeq (r_1, r_2)$, where $r_1$ and $r_2$ are irreducible representations of $\mathfrak{h}_1$ and $\mathfrak{h}_2$, respectively. In this case, the representation matrices of the generators act as
\begin{align}
\tau_{A_{(1)}} = T_{A_{(1)}}^{(r_1)} \otimes \mathbf{1}_{r_2}, \qquad \tau_{A_{(2)}} = \mathbf{1}_{r_1} \otimes T_{A_{(2)}}^{(r_2)}.
\end{align}
This structure preserves the commutation relations of each subalgebra. From the definition of $I_i$ as a trace over $\mathfrak{m}$, one immediately obtains the factorization rule:
\begin{align}
I_1(\mathfrak{m}) = I_1(r_1) \dim(r_2), \qquad I_2(\mathfrak{m}) = I_2(r_2) \dim(r_1).
\end{align}
This rule reduces the computation of $I_i$ for complex stabilizers to the indices of irreducible representations of their simple factors.

For example, in the cases with any factor \( \mathfrak{h}_i \simeq \mathfrak{so}(n) \), the relevant index ratios for the fundamental (\( \mathbf{n} \)), the trace-free symmetric tensor (\( \mathrm{Sym}^2_0(\mathbf{n}) \)), and the spinor (\( \mathbf{spin} \)) representations are given by:
\begin{equation}
   I_{\mathfrak{so}(n)}(\mathbf{n}) = \frac{1}{n-2}, \quad 
   I_{\mathfrak{so}(n)}(\mathrm{Sym}^2_0(\mathbf{n})) = \frac{n+2}{n-2}, \quad 
   I_{\mathfrak{so}(n)}(\mathbf{spin}) = \frac{\dim(\mathbf{spin})}{8(n-2)}.
\end{equation}
These formulas, combined with the factorization rule, allow for the systematic reproduction of the \( I_i \) values for many series in Table~\ref{tab:Ihvaluelist}. For example, in the BDI series where \( \mathfrak{m} = (\mathbf{p}, \mathbf{q}) \) under \( \mathfrak{h} = \mathfrak{so}(p) \oplus \mathfrak{so}(q) \), the index for \( \mathfrak{so}(p) \) is immediately obtained as \( I_1 = I_{\mathfrak{so}(p)}(\mathbf{p}) \dim(\mathbf{q}) = q/(p-2) \). Similarly, the results for the exceptional series involving spinor representations, such as EIII, EVIII and FII, are directly computed from these algebraic relations.

In practice,  we derived all values in Table~\ref{tab:Ihvaluelist} using the diagrammatic "birdtrack" method \cite{Cvitanovic:1976am,Cvitanovic:2008zz}.
This method allows for a direct evaluation of invariants by representing the projectors onto the irreducible components of the isotropy representation $\mathfrak{m}$ as diagrammatic contractions. This approach is particularly powerful for the exceptional series (e.g., $F_4, E_8$), where the index ratios can be computed systematically without the need for explicit weight-counting or referring to large tables of characters. For the classical series, the results are easily verified to be consistent with the well-known ratios of Dynkin indices for tensor representations.

\section{Five-term model: Robust AdS preference with $S^n$ and $\mathbb{C}P^m$}\label{sec:RobustAdS}

Let us consider a five-term model obtained by adding a quartic Lovelock term to the four-term model given in Eq.~\eqref{eq:4term}.
Under the assumption that a Minkowski vacuum exists, the potential can be written as
\begin{align}
V_0(\rho)
= \left( \frac{\rho}{\hat \rho_0} \right)^{\hat d - 8}
\left( \left( \frac{\rho}{\hat \rho_0} \right)^2 - 1 \right)^2
\, h\left( \left( \frac{\rho}{\hat \rho_0} \right)^2 \right),
\qquad (\hat d \ge 8),
\end{align}
where the polynomial function $h$ is defined by
\begin{align}
h(x) := c_0 x^2 + c_1 x + c_2,
\end{align}
with three parameters $c_{0,1,2}$.
By setting $c_2 = 0$, this model reduces to the four-term model.
This choice implies that the Lovelock coefficients are given by
\begin{align}
\hat \alpha_k
= -\frac{1}{\LL_k(\hat g)\,\hat \rho_0^{\hat d - 2k}}
\left( c_k - 2 c_{k-1} + c_{k-2} \right),
\qquad
c_m = 0 \quad \text{for } m\notin [0,2],
\end{align}
and hence $K(\hat \rho_0)$ takes the form
\begin{align}
K(\hat \rho_0)
= -\frac{\hat \rho_0^2}{R(\hat g)}
\left( \nu_0 c_0 + \nu_1 c_1 + \nu_2 c_2 \right),
\end{align}
where the coefficients $\nu_k$ are defined as
\begin{align}
\nu_k
:= k(\gamma_{k+1} + \gamma_{k-1} - 2\gamma_k)
+ 2(\gamma_{k+1} - \gamma_k),
\qquad
\gamma_m = 0 \quad \text{for } m<0.
\end{align}

\begin{figure}[h]
\centering
\includegraphics[width=0.6\textwidth,trim=0 100 0 100, clip]{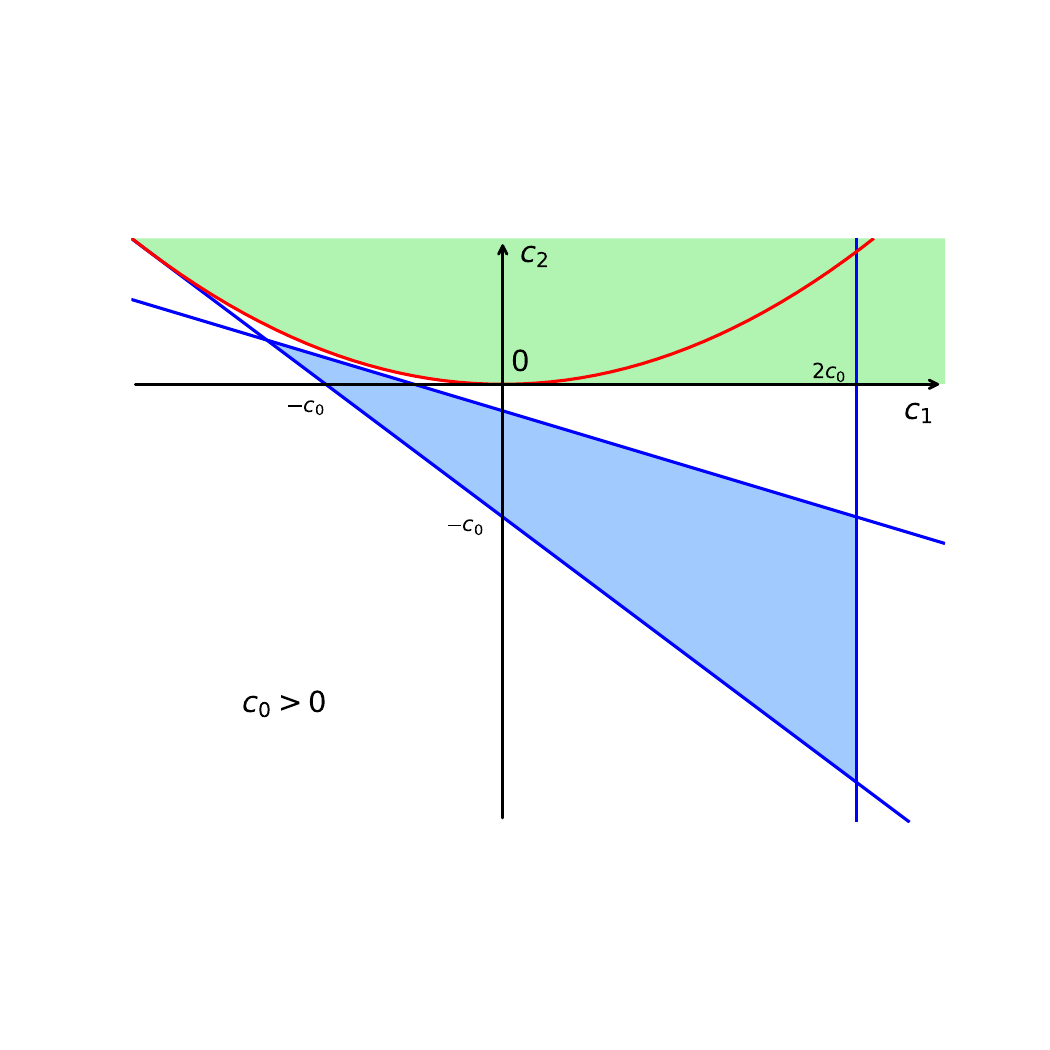}
\caption{Ghost-free and locally-stable region (light blue) and globally-stable region (light green) in the five-term model.}
\label{fig:RobustMetaStability}
\end{figure}

In this model, the three conditions for the Minkowski vacuum to be ghost-free and locally stable,
$\alpha_1 > 0$, $V''(\hat\rho_0) > 0$, and $K(\hat\rho_0) > 0$,
can be written as
\begin{align}
& 2c_0 - c_1 > 0,
\qquad
c_0 + c_1 + c_2 > 0,
\qquad
\nu_0 c_0 + \nu_1 c_1 + \nu_2 c_2 < 0.
\end{align}
If the coefficients $\nu_{0,1,2}$ satisfy
\begin{align}
\nu_2 > \nu_1 > \nu_0 > 0,
\label{eq:nu-convexity}
\end{align}
one can show that $c_0$ must be positive ($c_0>0$),
and the allowed parameter region becomes a triangular domain,
illustrated in Fig.~\ref{fig:RobustMetaStability} in light blue.

Global stability of the vacuum requires that the equation
$h(x)=0$ has no real roots for $x>0$,
which leads to the condition
\begin{align}
\left( 4 c_0 c_2 > c_1^2 \right)
\;\lor\;
\left( (c_1 \ge 0) \land (c_2 \ge 0) \right),
\end{align}
shown in Fig.~\ref{fig:RobustMetaStability} in light green.

Moreover, if the coefficients satisfy
\begin{align}
\nu_2 \nu_0 > \nu_1^2,
\label{eq:nu-log-convexity}
\end{align}
in addition to \eqref{eq:nu-convexity},
then the above two regions never overlap.

We find that for both $S^n$ and $\mathbb{C}P^m$
all inequalities in \eqref{eq:nu-convexity} and \eqref{eq:nu-log-convexity}
are satisfied for arbitrary $n\ge 8$ and $m\ge 4$
as follows.
First, we show $\nu_1>\nu_0>0$ for general CIRS spaces as
\begin{align}
    \nu_0=2(\gamma_1-1)>0,\quad \nu_1-\nu_0=3(\gamma_2-2\gamma_1+1)
    =3(1+\mu)(\gamma_1-1)^2>0.
\end{align}
Using the explicit formulas for $\gamma_k$ in \eqref{eq:gammaSnCPm},
Inequality \eqref{eq:nu-log-convexity} can be checked explicitly.
For $S^n$ case, setting $p=n-8\ge 0$ gives
\begin{align}
    &\nu_2\nu_0-\nu_1^2\nonumber\\
    &=\frac{192(5 p^6+189 p^5+ 2795 p^4 + 20847 p^3+ 82817 p^2 + 165543 p + 128682)}{(p + 1)(p+ 2)(p + 3)^2(p + 4)^2(p + 5)^2(p+ 6)^2}>0,
\end{align}
and for $\mathbb CP^m$ case with $m\ge 4$,
\begin{align}
   \nu_2\nu_0-\nu_1^2=\frac{12(5m + 17)}{(m - 3)(m - 2)^2(m - 1)^2}>0.
\end{align}
Furthermore, since $\nu_1>\nu_0>0$, Inequality \eqref{eq:nu-log-convexity} implies $\nu_2>\nu_1^2/\nu_0>\nu_1$.

This implies that, at least in these cases, the Minkowski vacuum is always metastable, and thus the robust AdS preference holds. Since \(S^n\) and \(\mathbb{C}P^m\) represent the cases where the log-convexity of \(\gamma_k\) is the ``weakest,'' we conjecture that Inequalities~\eqref{eq:nu-convexity} and \eqref{eq:nu-log-convexity}t aret satisfied generally, and thus the robust AdS preference persists.

It is remarkable that the convexity originating from the CIRS geometry governs the physics repeatedly across different hierarchical levels: first through the log-convexity of \(\gamma_k\) in the four-term model, and here through the higher-order convexity of \(\nu_k\) in the five-term model. This confirms that the AdS preference is not a mere artifact of model truncation, but a persistent structural consequence of the underlying algebraic framework.

\bibliographystyle{utphys}
\bibliography{references}

\end{document}